\documentclass[fleqn,usenatbib]{mnras}

\usepackage{lineno}
\usepackage{newtxtext,newtxmath}
\usepackage{multirow}
\usepackage[T1]{fontenc}

\DeclareRobustCommand{\VAN}[3]{#2}
\let\VANthebibliography\thebibliography
\def\thebibliography{\DeclareRobustCommand{\VAN}[3]{##3}\VANthebibliography}

\usepackage{graphicx}	
\usepackage{amsmath}	

\title[Evolution of CME Aspect Ratio]{Three-Phase Evolution of Aspect Ratio in Fast and Slow CMEs from the Sun to 1 AU}

\author[Mishra et al.]{
Wageesh Mishra$^{1,2}$\thanks{E-mail: m.wageesh30@gmail.com},
Anjali Agarwal$^{1,2}$\thanks{E-mail: anjaliagarwal1024@gmail.com}
and Nandita Srivastava$^{3}$
\\
$^{1}$Indian Institute of Astrophysics, II Block, Koramangala, Bengaluru 560034, India\\
$^{2}$Pondicherry University, R.V. Nagar, Kalapet 605014, Puducherry, India\\
$^{3}$Udaipur Solar Observatory, Physical Research Laboratory, Udaipur 313001, India
}

\date{Accepted 2026 June 10. Received 2026 May 19; in original form 2025 December 26}

\pubyear{\the\year{}}

\begin{document}

\label{firstpage}
\pagerange{\pageref{firstpage}--\pageref{lastpage}}
\maketitle

\begin{abstract}
Coronal mass ejections (CMEs) undergo significant geometric evolution as they propagate from the Sun to 1 AU, influencing their radial size, expansion, and space weather impact. We investigate the evolution of CME aspect ratio ($\kappa$) and expansion dynamics for four fast and four slow Earth-directed CMEs. Using multipoint coronagraphic observations with the Graduated Cylindrical Shell (GCS) model and corrected in situ measurements of associated magnetic clouds (MCs) at 1 AU, we track the evolution of $\kappa$ from the low-middle corona to interplanetary space. We find that $\kappa$ does not remain constant but exhibits a systematic three-phase evolution: a rise phase in the low-middle corona ($\lesssim10$--$15\,R_{\odot}$), a saturation phase at intermediate heights, and then a decline phase in the interplanetary space. The ratio of radial expansion speed to leading-edge speed ($V_{\rm exp}/V_{\rm LE}$) decreases substantially from the corona to 1 AU, indicating a reduction in radial expansion efficiency during interplanetary propagation. The consistent evolution of $\kappa$ and $V_{\rm exp}/V_{\rm LE}$ suggests a transition from magnetically dominated expansion in the corona to a regime increasingly controlled by the heliospheric environment. We note that fast CMEs show stronger early expansion and evolve into larger, more radially extended structures, whereas slow CMEs exhibit a more gradual rise and a steeper decline. These results demonstrate that CME geometry evolves significantly during propagation and highlight the need to incorporate aspect ratio evolution in models to improve predictions of CME size, arrival time, and geoeffectiveness.
\end{abstract}


\section{Introduction}\label{sec:intro}

Coronal Mass Ejections (CMEs) are the episodic expulsion of billions of tons of magnetized plasma from the solar corona into interplanetary space. These large-scale eruptions on their arrival to Earth have the potential to drive significant space weather events near Earth, causing risks to technological infrastructure and human activities in space \citep{Schwenn2006,Webb2012,}. The studies on CME behavior in the interplanetary space have largely attempted to track the CMEs away from the Sun, understand their dynamics under different forces acting on it, develop their arrival time models, investigate the nature of interacting CMEs, and the evolution in the characteristics of its substructures (shock, sheath, flux rope, leading and trailing edges) \citep{Gopalswamy2000,Owens2005, Davies2009,Mishra2013,Mostl2014,Mishra2017,Harrison2018,Heinemann2019,Barnard2023,Khuntia2025,Turner2025,Agarwal2026}. These studies have integrated observations from multiple spacecraft, capturing CMEs through both remote sensing and in situ measurements alongside modeling efforts. Although a few decades of research in CMEs have provided key insights about their initiation, evolution, and impacts on planetary environments \citep{Hundhausen1999,Schrijver2003,Mishra2015,Manchester2017,Temmer2023}, but still there remains a large gap in understanding the geometrical and dynamic evolution of CMEs.

Regarding estimating the deprojected kinematics of CMEs, i.e., the true radial motion in 3D obtained by correcting projection effects present in plane-of-sky observations, there has been a plethora of studies utilizing a variety of reconstruction methods suitable for coronagraphic and heliospheric imaging observations of CMEs \citep{Davies2009,Thompson2009,Mierla2010,Davies2013,Mishra2014,Mishra2023}. Most of these methods and their implementation assume a constant CME geometry and attempt to model the dynamics of CMEs. Several studies have incorporated deformation of CME geometry to better represent its evolving structure and have shown that such geometrical evolution provides insights into CME expansion dynamics, interaction with the ambient solar wind, and the mechanisms governing their interplanetary evolution  \citep{Owens2006a,Savani2011,Kay2021,Isavnin2016,Hinterreiter2021,Braga2022,Barnard2023}. The Graduated Cylindrical Shell (GCS) model \citep{Thernisien2009} of CMEs has the potential to provide the geometrical properties of the CMEs' flux rope, but the majority of studies have utilized it to understand only the kinematic evolution of CMEs.

The GCS model for a CME involves estimating its aspect ratio, an important geometric parameter that characterizes CME evolution. In this model, the aspect ratio is defined as the ratio of the CME (flux rope) radius to the height of the CME (flux rope) center from the Sun \citep{Thernisien2011}. The aspect ratio affects both the radial and lateral dimensions of CMEs and plays a crucial role in understanding their expansion dynamics, thermodynamics, and arrival times \citep{Savani2011,Mishra2018,Mishra2023a,Agarwal2024}. Although CMEs appear to have a continuous change in their aspect ratio, there are limited studies in this direction to examine the evolution of aspect ratio as CMEs evolve in the corona \citep{Cremades2020,Agarwal2024}.

In the literature, another commonly used approach to constrain the geometric (radial and lateral) evolution of CMEs is to assume a Self-Similar Expansion (SSE) of a CME. Under the SSE assumption for a CME, its circular cross-section expands/contracts uniformly with time while maintaining a fixed angular width with respect to the Sun’s center \citep{Davies2012,Subramanian2014,Demoulin2020}. This is possible when every plasma parcel maintains a constant fractional distance from the CME center relative to the radius of the cross-section, so that the internal spatial configuration of the CME is preserved while its size changes. Such an assumption of SSE during the CME propagation also implies that the aspect ratio of the CME flux rope remains constant. This assumption of constant aspect ratio (or SSE assumption) can introduce significant uncertainties in estimating CME kinematics \citep{Patsourakos2010,Barnard2017,Nieves-Chinchilla2018,Dumbovic2026}, and may lead to larger errors in key derived parameters such as arrival time, speed, flux-rope size, and magnetic flux. Therefore, attempts should be made to estimate the aspect ratio of varieties of CMEs and examine their constancy during the propagation of a CME.

It is worth noting that the aspect ratio of a CME is defined differently across the literature, and different approaches are adopted to estimate its value \citep{Krall2001,Patsourakos2010,Rollett2016,Veronig2018,Braga2020,Cremades2020}. Interestingly, even within the same modeling framework, studies have used inconsistent definitions of the aspect ratio. For instance, \citet{Krall2001} and \citet{Veronig2018} both employed an elliptical fitting approach to the CME’s projected image, but adopted different definitions, whereas \citet{Patsourakos2010} and \citet{Cremades2020}, despite using the GCS model, also defined the aspect ratio differently. The studies of \citet{Patsourakos2010} and \citet{Veronig2018} have considered the aspect ratio as the ratio of CME flux rope center height from the Sun to its radius, whereas other studies \citep{Krall2001,Krall2006,Cremades2020} have considered it as the ratio of CME flux rope center height to its diameter. In the Ellipse Evolution Model (ElEvo), the aspect ratio refers to the ratio between its lateral (perpendicular to propagation direction) dimension and its radial (along the propagation direction) dimension \citep{Savani2011,Mostl2015,Rollett2016,Davies2021,Zhuang2022}.In such studies, a CME with a higher aspect ratio is more elongated in the direction perpendicular to its propagation. However, despite using the elliptical front model for CMEs, \citet{Braga2020} has considered the aspect ratio as the ratio of the CME radial dimension to its lateral dimension.

Different definitions of the aspect ratio have led to a varying range of its values reported in the literature \citep{Krall2001,Thernisien2009,Patsourakos2010,Rollett2016,Veronig2018,Braga2020,Cremades2020,Agarwal2024}. This has made it challenging to compare aspect ratio estimates across different studies, leading to variations in the interpretation of CME morphology and expansion properties. It has occasionally led to incorrect comparisons of estimates, as seen in \citet{Braga2020}, who compared the values obtained with those of \citet{Rollett2016}. Furthermore, some studies have used the GCS model-derived aspect ratio to estimate geometric parameters (such as height, radius, and speed), but considered different definitions of the aspect ratio from that originally defined in the GCS framework. \citep{Patsourakos2010,Cremades2020}. Therefore, it is imperative to accurately calculate the aspect ratio of CMEs near the Sun from the GCS model for several CMEs and understand their trends for different types of CMEs.

Another limitation of prior studies is that most of them have estimated the value of the aspect ratio within the coronagraphic field of view only. However, the aspect ratio of CMEs evolves as they propagate outward from the Sun due to various physical processes, including internal expansion, interaction with the ambient solar wind, and the influence of background magnetic fields. Moreover, differences in the evolution of aspect ratio between fast and slow CMEs can provide insights into the kinematic and dynamic forces acting on these structures. Fast CMEs, typically associated with strong magnetic fields and higher kinetic energies, may tend to expand more rapidly, while slow CMEs evolving more gradually may experience stronger confinement and coupling by the ambient solar wind \citep{Sachdeva2017}. It is essential to determine whether slow and fast CMEs exhibit distinct evolutionary trends in their aspect ratio from the Sun to Earth, as such differences reflect their underlying expansion dynamics, interaction with the ambient solar wind, and ultimately influence their propagation and geoeffectiveness. Yet, systematic studies on their geometrical evolution remain sparse.

In this study, we take the aspect ratio as the ratio of the CME flux-rope radius to the height of the CME center \citep{Thernisien2009,Thernisien2011}. The aspect ratio near the Sun is obtained from the GCS model applied to multi-viewpoint coronagraph observations (Section~\ref{sec:gcsaspect}), whereas the aspect ratio at 1 AU is derived from the flux-rope radius estimated from in situ measurements (Section~\ref{sec:aspectmc1au}). Although these values are obtained using different techniques on coronagraph and in situ observations, both quantify the aspect ratio as the ratio of flux rope radius to the height of the flux rope center from the Sun. They therefore rely on consistent geometric definitions of the CME flux-rope cross-section, allowing a meaningful comparison of its evolution from the Sun to 1 AU. By analyzing the temporal evolution of the aspect ratio for both fast and slow CMEs from coronal heights to 1 AU, we investigate their evolutionary profiles and associated expansion behavior. In particular, we aim to identify systematic differences in the radial growth and expansion dynamics of slow and fast CMEs. This study provides insight into the physical mechanisms governing the evolution of CME aspect ratio and has important implications for improving space weather forecasting.

\section{OBSERVATIONS OF SELECTED CMES AND ANALYSIS}{\label{sec:observation}}

\begin{figure*}
    \centering
    \includegraphics[scale= 0.4,trim={0cm 0cm 0cm 0cm},clip]{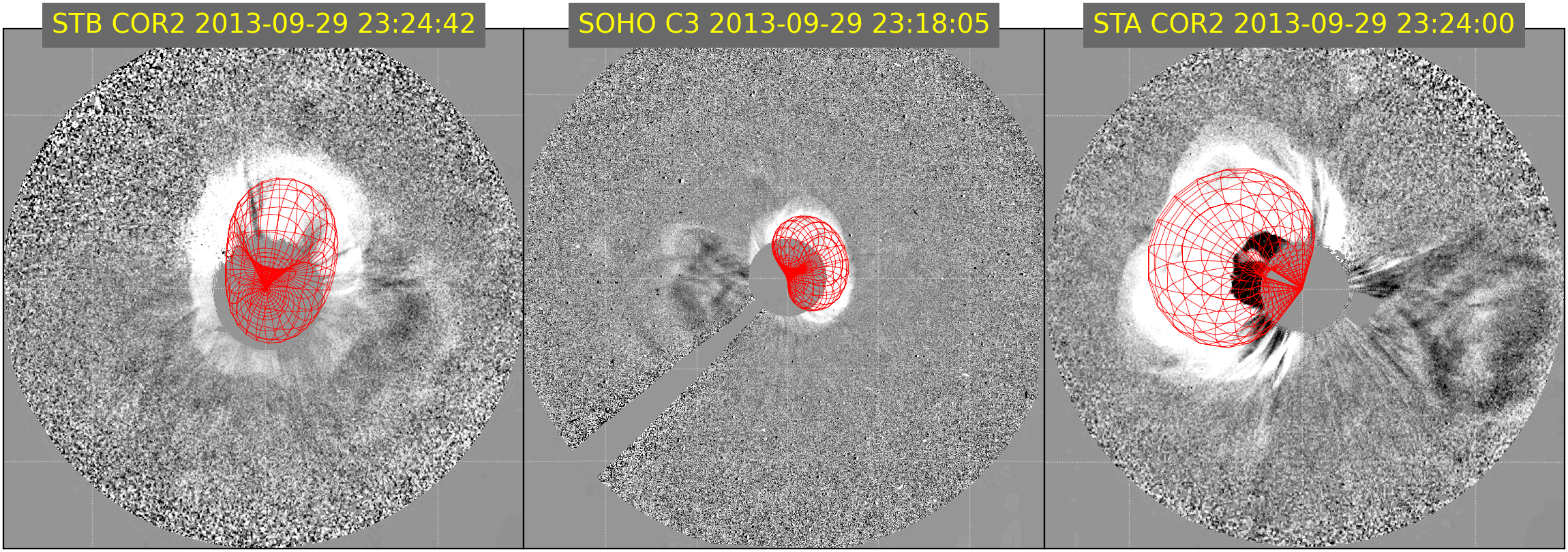}
   \includegraphics[scale=0.4,trim={0cm 0cm 0cm 0cm},clip]{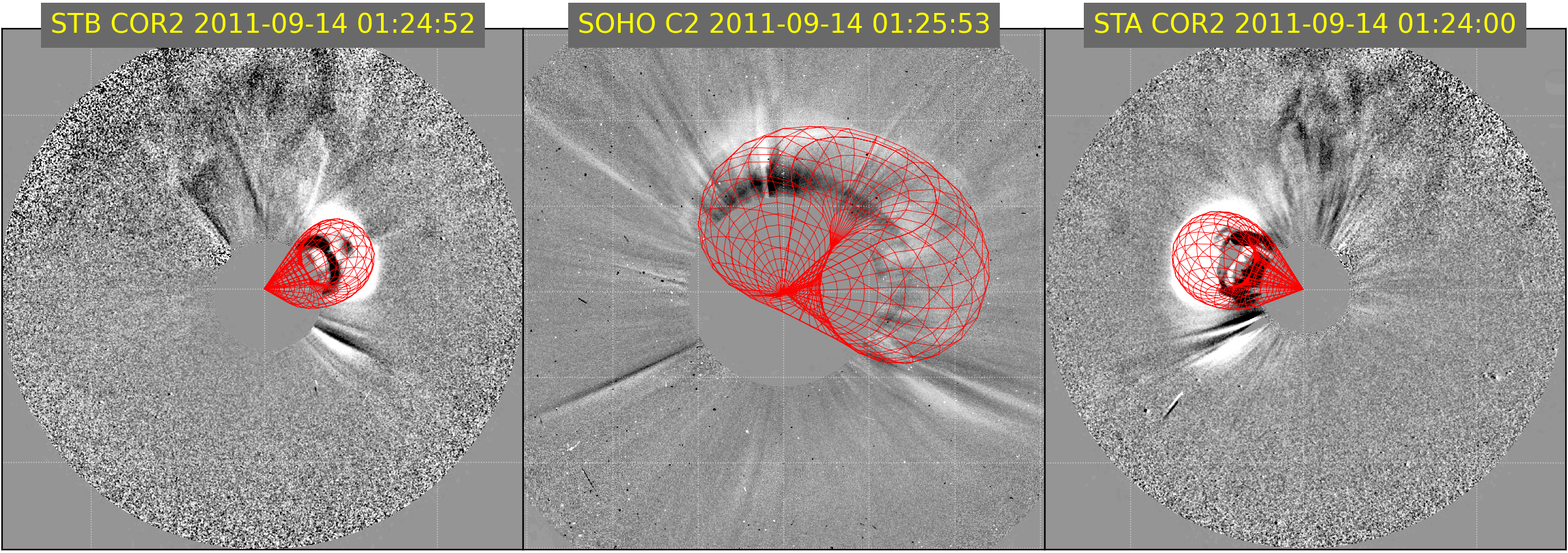}
    \caption{The top and bottom panels show the GCS model fitting of fast and slow CMEs of 2013 Sep 29, and 2011 Sep 13, respectively. The model is implemented on the contemporaneous coronagraphic (COR2/C2/C3) observations from three viewpoints: COR2 on STB (left), C2/C3 on SOHO (center), and COR2 on STA (right). STB and STA represent the \textit{STEREO-B} and \textit{STEREO-A} spacecraft, respectively.}
    \label{fig:gcs}
\end{figure*}

We investigate the temporal evolution of the aspect ratio of both fast and slow CMEs from their initial coronagraphic heights to their arrival at 1 AU. For this study, we selected CMEs that met several stringent criteria, ensuring that the evolution of their aspect ratio could be accurately determined in both coronagraphic observations near the Sun and in situ observations at 1 AU. These criteria are: (i) CMEs were chosen from the \textit{STEREO} era from late 2008 to late 2013, during which the angular separation between \textit{SOHO} and the twin \textit{STEREO} spacecraft ranged from around 40$^\circ$ to 145$^\circ$, allowing for reliable 3D reconstructions of CMEs flux rope's morphology \citep{Lyu2021,Lyu2023} (ii) we ensured that the selected CMEs propagated as isolated events without significant interactions with other preceding CMEs so that their measured morphology near the Sun and the in situ measured plasma parameters near 1 AU would both be representative of isolated CMEs (iii) we also ensured that selected CMEs arrived on the Earth within a longitude range of $\pm$30$^\circ$ from the Sun-Earth line based on the GCS model (iv) we further ensure that selected CMEs on their arrival to Earth are classified as magnetic clouds (MCs), a subset of ICMEs \citep{Zurbuchen2006}, in situ observations to ensure reliable estimates of their dimension and aspect ratio at 1 AU.

Based on the above-mentioned criteria, we selected front-side halo CMEs from the CDAW catalog (\url{https://cdaw.gsfc.nasa.gov/CME_list/halo/halo.html}) and analyzed coronagraphic observations from both \textit{SOHO} and \textit{STEREO}. We estimated CMEs' true direction of propagation using the GCS model and verified their arrival as MCs at Earth from Richardson and Cane ICME catalog (\url{https://izw1.caltech.edu/ACE/ASC/DATA/level3/icmetable2.htm}), where MCs are defined as structures exhibiting enhanced magnetic field strength ($>$10 nT), smooth rotation of the magnetic field vector through a large angle (typically $\gtrsim 30^\circ$), low proton temperature, and low plasma beta ($<$1) \citep{Lepping1990,Huttunen2005}. Based on these criteria, we selected a set of eight CMEs for our study, categorizing 4 as fast (CMEs of 2011 Jun 02, 2012 Jul 12, 2013 Apr 11 and 2013 Sep 29)  and 4 as slow (CMEs of 2008 Dec 12, 2010 Apr 03, 2010 Apr 08 and 2011 Sep 13), as listed in Table~\ref{tab:GCS_para}. All of the selected CMEs have been have been previously analyzed \citep{Isavnin2013,Mishra2014,Shen2014,Zhuang2017,Sachdeva2017,Temmer2021}, most of earlier studies did not specifically investigate the temporal evolution of the aspect ratio, as the GCS model is commonly applied under the assumption of self-similar expansion with a constant aspect ratio.

In this study, CMEs are classified as fast or slow based on the average speed of the leading edge (LE), defined as the outermost bright front observed in coronagraph images. This average is estimated from the 3D LE speeds measurements at the first three tracked heights using the GCS model in COR1 for slow CMEs (except the 2008 Dec 12 event) and in COR2 for fast CMEs (except on 12 Jul 2012 and 11 Apr 2013). A threshold of 600 km s$^{-1}$ (considered at the first three tracked heights for each CME), representing the typical average between fast and slow solar wind speeds, is used to compare with the estimated average LE speeds--CMEs exceeding this value are classified as fast, while those below it are classified as slow. The 3D speeds are derived using the GCS model \citep{Thernisien2009}, which enables a 3D reconstruction of CMEs from multi-viewpoint coronagraphic observations, as described in the following section.

\subsection{GCS Model Fitting and Aspect Ratio of CMEs at Coronal Heights}
\label{sec:gcsaspect}

To estimate the evolution of the aspect ratio of CMEs near the Sun, we employed the GCS model \citep{Thernisien2009} to contemporaneous coronagraphic images obtained from \textit{SOHO} and the twin \textit{STEREO} spacecraft \citep{Brueckner1995,Howard2008,Kaiser2008}. The model approximates the CME flux rope as a hollow croissant, parameterized by six free parameters: latitude ($\theta$), longitude ($\phi$), half-angle between conical legs ($\alpha$), tilt angle ($\gamma$), aspect ratio ($\kappa$), and the height ($h$) of the CME LE. Figure~\ref{fig:gcs} presents the GCS model fittings for a fast CME launched on 2013 Sep 29 (top panel) and a slow CME launched on 2011 Sep 13 (bottom panel), using coronagraphs onboard \textit{SOHO}, \textit{STEREO-A} and \textit{STEREO-B}, illustrating their morphology from three different viewpoints.

\begin{table*}
    \centering
    \footnotesize
    \begin{tabular}{ccccccc}
    \hline
    {Date} & h ($R_\odot$) & {$\theta~(^\circ)$} & $\phi~(^\circ)$ & $\alpha~(^\circ)$ & $\gamma~(^\circ)$ & $\kappa$ \\
    
     & ($\pm$ MAD) & (median $\pm$ MAD) & (median $\pm$ MAD) & (median $\pm$ MAD) & (median $\pm$ MAD) & (median $\pm$ MAD)\\
     &  [(MAD/h)\%] &  &  & [(MAD/median)\%] &  & [(MAD/median)\%]\\
    \hline
    \multicolumn{7}{c}{Fast CMEs} \\
    \hline
    2011 Jun 02 & 13.5 ($\pm$ 1.4) [$\sim$10] & -10 (-4$\pm$3) & -15 (-28$\pm$3) & 20 (33$\pm$9) [$\sim$30] & 50 (27$\pm$19) & 0.23 (0.35$\pm$0.05) [$\sim$15]\\
    
    2012 Jul 12 & 14.9 ($\pm$ 1.5) [$\sim$10] & -14 (-13$\pm$2) & 6 (2$\pm$4) & 25 (55$\pm$14) [$\sim$25] & 40 (78$\pm$12) & 0.61 (0.48$\pm$0.03) [$\sim$5]\\
    
    2013 Apr 11 & 16.3 ($\pm$ 2.4) [$\sim$15] & 0 (-1$\pm$6) & -20 (-11$\pm$5) & 35 (45$\pm$14) [$\sim$30] & 60 (78$\pm$18) & 0.40 (0.29$\pm$0.04) [$\sim$15]\\
    
    2013 Sep 29 & 15.2 ($\pm$ 1.5) [$\sim$10] & 25 (25$\pm$3) & 30 (25$\pm$4) & 40 (49$\pm$9) [$\sim$20] & -80 (90$\pm$7) & 0.60 (0.43$\pm$0.05) [$\sim$10] \\
    \hline
    \multicolumn{7}{c}{Slow CMEs} \\
    \hline
    2008 Dec 12 & 15.1 ($\pm$ 1.5) [$\sim$10] & 10 (7$\pm$1) & 15 (7$\pm$2) & 20 (23$\pm$5) [$\sim$20] & 55 (38$\pm$25) & 0.23 (0.27$\pm$0.01) [$\sim$5]\\
    
    2010 Apr 03 & 13.7 ($\pm$ 0.7) [$\sim$5] & -24 (-25$\pm$1) & 3 (4$\pm$3) & 25.5 (30$\pm$5) [$\sim$15] & 9.8 (2$\pm$8) & 0.37 (0.34$\pm$0.05) [$\sim$15]\\
    
    2010 Apr 08 & 13.8 ($\pm$ 0.7) [$\sim$5] & -5 (-5$\pm$4) & -5 (1$\pm$6) & 50 (27$\pm$6) [$\sim$20] & -30 (-18$\pm$11) & 0.28 (0.23$\pm$0.03) [$\sim$15]\\
    
    2011 Sep 13 & 14.1 ($\pm$ 0.7) [$\sim$5] & 20 (20$\pm$1) & 20 (18$\pm$3) & 25 (39$\pm$10) [$\sim$25] & -30 (-37$\pm$21) & 0.43 (0.38$\pm$0.06) [$\sim$15]\\
    \hline 
    \end{tabular}
    \caption{GCS-fitted CME parameters and associated uncertainties for the selected fast (top panel) and slow (bottom panel) CMEs at their last tracked coronal heights. The columns list the date, height ($h$), latitude ($\theta$), longitude ($\phi$), half-angle ($\alpha$), tilt angle ($\gamma$), and aspect ratio ($\kappa$). Uncertainties in $\theta$, $\phi$, $\alpha$, $\gamma$, and $\kappa$ are based on the median and median absolute deviation (MAD) from the LLAMACoRe catalog (see Section~2.2). Height uncertainties are estimated from the fractional MAD in CME speed (i.e., MAD/v) and assumed to scale with height. Percentage uncertainties are also indicated for height, half-angular width, and aspect ratio.}
    \label{tab:GCS_para}
\end{table*}

These events are well observed, can be reliably tracked throughout the coronagraphic field of view, and have simultaneous three-viewpoint observations available, thereby minimizing projection effects and making them well suited for analysis. Therefore, we were able to track the CMEs with reasonable precision (within the typical uncertainties as discussed in Section~\ref{sec:errorsgcs}) and examine the evolution of their 3D-fitted parameters from heights in the middle corona (1.5--6~$R_\odot$) to heights in the outer corona. The GCS parameters at the last tracked height are listed in Table~\ref{tab:GCS_para}, with fast CMEs shown in the top panel and slow CMEs in the bottom panel. The corresponding uncertainties in the GCS parameters are also provided in the table.  The listed uncertainties are derived from the reported median and MAD estimates in the LLAMACoRe (Living List of Attributes Measured in Any Coronal Reconstruction; \citet{Kay2024}) catalog. Details of the error estimation are described in Section~\ref{sec:errorsgcs}. We find that our GCS-fitted parameters are consistent with the catalog best-constrained values within the corresponding median absolute deviation (MAD) ranges and the uncertainties reported in earlier studies \citep{Thernisien2009,Pluta2019,Verbeke2023,Nikou2025}, supporting the reliability of our reconstructions for the selected well-observed CMEs.

From the table, we note that in general, fast CMEs tend to exhibit larger half-angular widths ($\alpha$) and aspect ratios ($\kappa$) compared to slow CMEs, indicating a greater degree of radial expansion. The final tracked heights are comparable across both categories, suggesting that the observed differences in $\alpha$ and $\kappa$ are primarily due to intrinsic expansion behavior rather than differences in the observational range. The propagation directions ($\theta$ and $\phi$) show that the selected CMEs are Earth-directed, with only a minimal offset from the equatorial plane. The tilt angles ($\gamma$) display significant event-to-event variability, reflecting diverse flux-rope orientations rather than a speed-dependent trend. The associated uncertainties in each GCS model parameter indicate that our GCS model fitting is reasonably well constrained for both CME groups.

\begin{figure*}
    \centering
    \includegraphics[scale= 0.72,trim={0cm 0cm 0cm 0cm},clip]{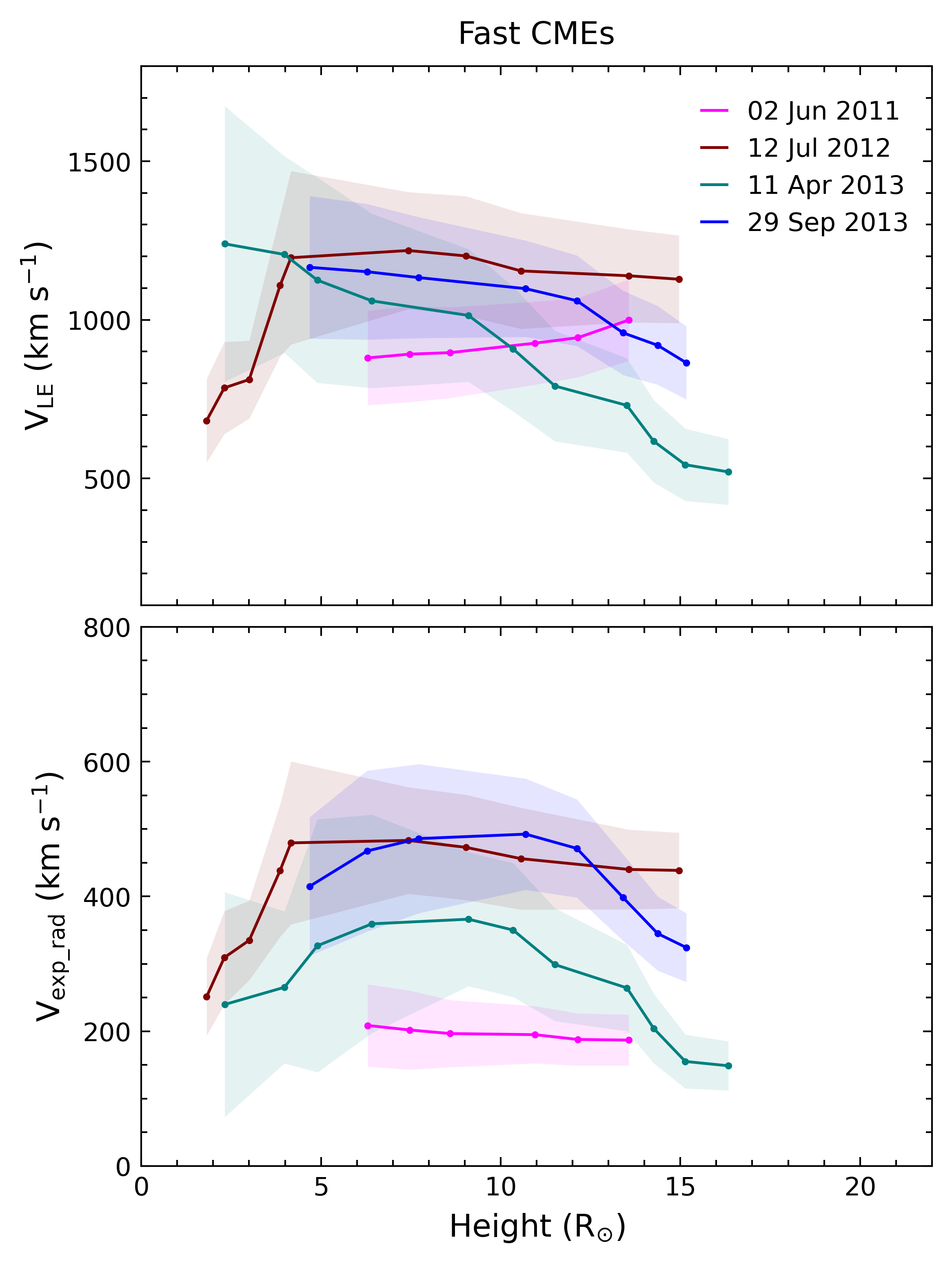}
    \includegraphics[scale=0.72,trim={0cm 0cm 0cm 0cm},clip]{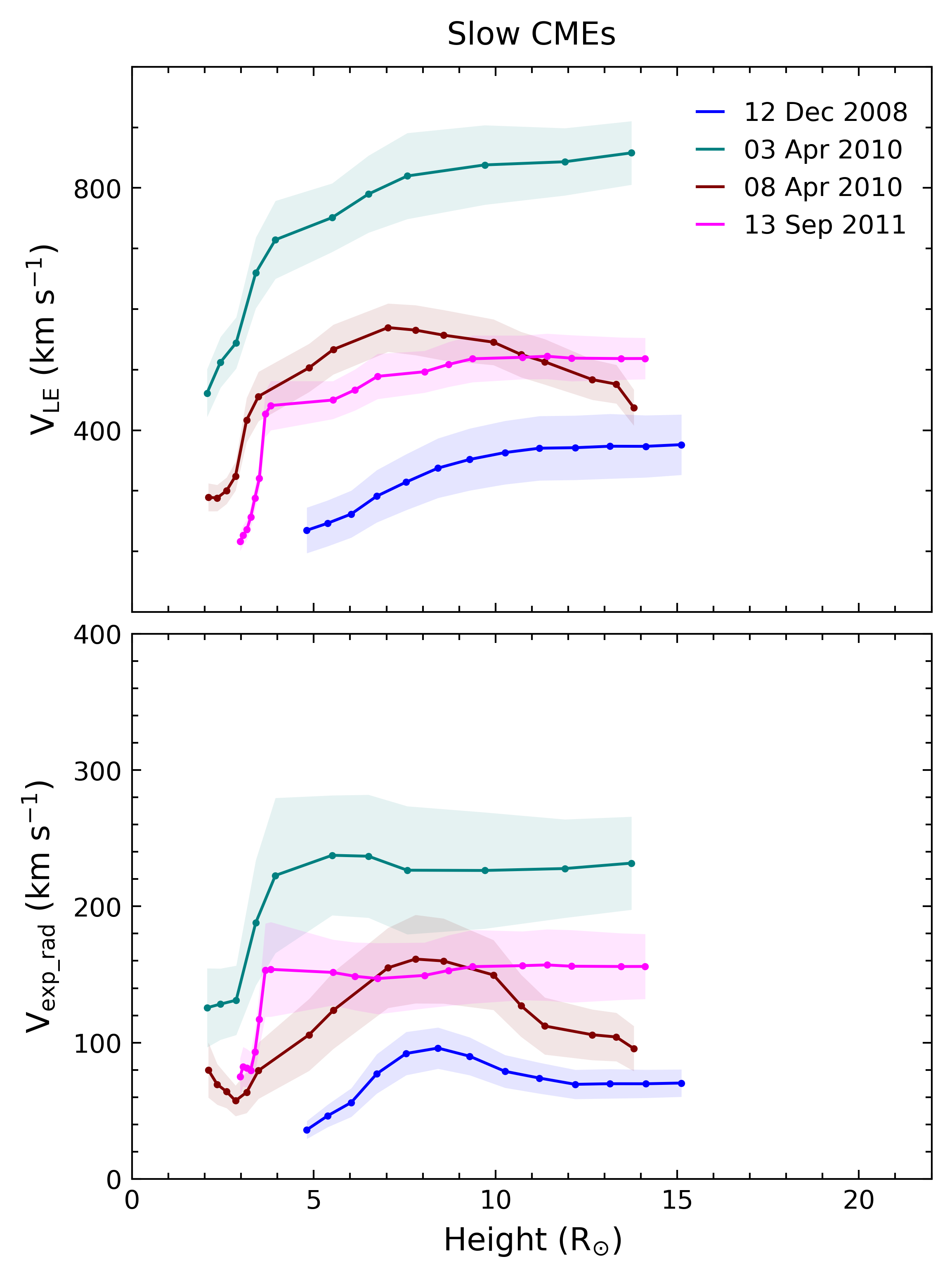}
    \caption{Height evolution of leading-edge speed ($V_{\rm LE}$; top panels) and radial expansion speed ($V_{\mathrm{exp\_rad}}$; bottom panels) for the selected fast (left panels) and slow (right panels) CMEs. The dots over the solid lines represent estimates derived from the GCS fitted values, while the shaded regions indicate the associated uncertainties derived from fractional uncertainties in leading edge height, radius, and aspect ratio of CME, as described in Section~\ref{sec:errorsgcs} and listed in Table~\ref{tab:GCS_para}.}
    \label{fig:le_exp_kin}
\end{figure*}

We estimated the 3D speed of the CME LE using a moving-box linear fit technique applied to height-time measurements from the GCS model \citep{Agarwal2024}. Figure~\ref{fig:le_exp_kin} shows the variation of LE speed ($V_{\rm LE}$; top panels) and radial expansion speed ($V_{\mathrm{exp\_rad}}$; bottom panels) as a function of heliocentric distance for the selected fast (left panels) and slow (right panels) CMEs. In general, fast CMEs exhibit significantly higher $V_{\rm LE}$ compared to slow CMEs and show diverse kinematic profiles, including both acceleration and deceleration phases within the coronagraphic field of view. The legends are arranged chronologically by CME occurrence date. The error bars are derived taking a certain percentage error (as described in Section~\ref{sec:errorsgcs} and listed in the second column of Table~\ref{tab:GCS_para}) in the measurements of the height from the GCS model at each data point and represented by transparent fill areas over the data point with the same color as used for the data points. In contrast to fast CMEs, slow CMEs display more gradual and monotonic evolution in $V_{\rm LE}$, with modest acceleration or near-constant speeds, indicating weaker interaction with the ambient solar wind.

The radius of the CME flux rope is estimated using the well-established relations \citep{Thernisien2009,Thernisien2011}. In the GCS geometry, the CME flux rope is represented by a croissant-like structure, where the aspect ratio ($\kappa$) determines the ratio between the radius of the flux rope and the distance of the flux-rope center from the Sun. The CME flux-rope radius ($R$) can therefore be expressed in terms of the $\kappa$ and the CME flux rope leading-edge  (LE) height ($h$) from the Sun, as follows:

\begin{equation}\label{equ:1}
    R = (\frac{\kappa}{1 + \kappa})h
\end{equation}

This definition has been extensively used in earlier studies \citep{Mishra2018,Mishra2020,Khuntia2024,Agarwal2024}. The bottom panel of the Figure~\ref{fig:le_exp_kin} depicts the radial expansion speed for the fast CMEs (left panel) and slow CMEs (right panel) determined using a moving-box linear fit applied to the radius-time evolution of CMEs. From this figure, we note that the radial expansion speed profile differs from the LE speed profile for most CMEs, except for three CMEs on 2010 Apr 3, 2011 Sep 13, and 2012 Jul 12. The radial expansion speed of the 2010 Apr 8 and 2008 Dec 12 CMEs initially increases and then decreases, whereas their LE speed first increases and subsequently attains a near-constant value. Similarly, the radial expansion speeds of the 2013 Apr 11 and 2013 Sep 29 CMEs show a pattern of an initial increase followed by a decline, while their LE speeds continuously decelerate within the tracked heights. The CME of 2011 Jun 2 exhibits an almost constant expansion speed while its LE speed increases. We note that the LE speed represents the propagation speed of the outermost front of the CME, whereas the radial expansion speed quantifies the temporal increase in the CME flux-rope radius. Since the CME radius depends on both the leading-edge height and the aspect ratio, the radial expansion speed reflects the combined evolution of CME propagation and geometrical expansion.

For fast events, $V_{\mathrm{exp\_rad}}$ generally increases during the early stages and either saturates or decreases at larger heights, reflecting changes in internal/external pressure during propagation. In contrast, slow CMEs exhibit relatively steady, lower expansion speeds with only modest variations throughout the observed height range. Overall, the figure highlights distinct kinematic and expansion characteristics between fast and slow CMEs, with fast CMEs showing stronger dynamical evolution in both propagation and expansion.

Based on our GCS model fit, we find that most 3D CME parameters (except $\kappa$) exhibit minimal variation with increasing height. The exceptions are the CMEs of 2010 Apr 8 for which $\theta$ varies from 15$^\circ$ to -5$^\circ$ (with uncertainties of $\pm$4$^\circ$), which could be attributed to the CME's deflection. The second and third columns of Table~\ref{tab:tab_2} list the range of GCS-derived aspect ratio ($\kappa$) and leading-edge speed ($V_{\rm LE}$) from the initial to the final tracked coronal heights, along with their associated uncertainties (errors) for the selected fast and slow CMEs. The details of the error estimation are described in Section~\ref{sec:errorsgcs}. The variation in $\kappa$ at coronal heights reflects the dynamic nature of CME evolution.

From the range of $\kappa$ values in Table~\ref{tab:tab_2}, we note that at the last tracked coronal height, the upper-bound aspect ratios for fast CMEs are systematically higher (range $\sim0.23$$\pm0.03$--$0.60$$\pm0.06$) than those for slow CMEs (range $\sim0.23$$\pm0.01$--$0.43$$\pm0.06$). This indicates that fast CMEs tend to attain more radially extended geometries by the end of the coronagraphic field of view. Fast CMEs, characterized by higher leading-edge speeds, exhibit a broader range of $\kappa$ values and more rapid changes compared to slow CMEs, indicating stronger and more dynamic expansion in the corona. The fast CMEs exhibit larger variations, with $\Delta\kappa \sim 0.04$--$0.30$ (corresponding to $\sim20\pm10$\%--$300\pm100$\%), whereas slow CMEs show more moderate changes of $\Delta\kappa \sim 0.09$--$0.14$ (corresponding to $\sim25\pm10$\%--$100\pm30$\%), indicating comparatively weaker and more gradual expansion. A pronounced increase in $\kappa$ is observed for the fast CMEs of 2012 Jul 12 and 2013 Apr 11, with the former undergoing acceleration and the latter deceleration. A moderate increase is also seen for the decelerating fast CME of 2013 Sep 29. In contrast, all slow CMEs display only modest variations in their GCS-derived aspect ratios.

\begin{table*}
    \centering
    \setlength{\tabcolsep}{4pt}
    \footnotesize
    \begin{tabular}{cccccccccccccc}
    \hline
    \multirow{3}{2em}{Date} & \multicolumn{2}{c}{Range of $\kappa$} & \multicolumn{2}{c}{Range of $V_{LE}$} & \multicolumn{3}{c}{In situ measured} & \multicolumn{6}{c}{$V_{exp}/V_{LE}$ (\%)} \\
    \cline{9-14}
    & \multicolumn{2}{c}{in COR} & \multicolumn{2}{c}{(km s$^{-1}$) in COR } & \multicolumn{3}{c}{$\kappa$ at 1 AU} & \multicolumn{2}{c}{At final $h$} & \multicolumn{2}{c}{In situ measured} & \multicolumn{2}{c}{From equ.~\ref{equ:vexp/vle}} \\
    & \multicolumn{2}{c}{(Error)} & \multicolumn{2}{c}{(Error)} & \multicolumn{3}{c}{[Correction (abs.,\% )]} &  \multicolumn{2}{c}{in COR (Error)} & \multicolumn{2}{c}{at 1 AU (Correction)} & \multicolumn{2}{c}{at 1 AU (Error)} \\
    \hline
    \multicolumn{14}{c}{Fast CMEs} \\
    \hline
    2011 Jun 02 & \multicolumn{2}{c}{0.19 $(\pm$0.03) -- 0.23 ($\pm$0.03)} & \multicolumn{2}{c}{880 ($\pm$150) -- 1000 ($\pm$130) } & \multicolumn{3}{c}{0.11 [$+$0.024, 21.8]} & \multicolumn{2}{c}{18.7 ($\pm$4.5)} & \multicolumn{2}{c}{5.7 ($-$5.3)} & \multicolumn{2}{c}{6.7 ($\pm$2.4)} \\
    
    2012 Jul 12 & \multicolumn{2}{c}{0.36 ($\pm$0.02) -- 0.61 ($\pm$0.03)} & \multicolumn{2}{c}{682 ($\pm$130) -- 1128 ($\pm$140)} & \multicolumn{3}{c}{0.37 [$+$0.02, 5.4]} & \multicolumn{2}{c}{38.9 ($\pm$7)} & \multicolumn{2}{c}{15 ($-$2.7)} & \multicolumn{2}{c}{21.9 ($\pm$1.3)} \\
    
    2013 Apr 11 & \multicolumn{2}{c}{0.10 ($\pm$0.02) -- 0.40 ($\pm$0.07)} & \multicolumn{2}{c}{1240 ($\pm$430) -- 520 ($\pm$100)} & \multicolumn{3}{c}{0.14 [$+$0.015, 10.7]} & \multicolumn{2}{c}{28.6 ($\pm$9)} & \multicolumn{2}{c}{8.5 ($-$4.7)} & \multicolumn{2}{c}{6.5 ($\pm$1.6)} \\
    
    2013 Sep 29 & \multicolumn{2}{c}{0.42 ($\pm$0.04) -- 0.60 ($\pm$0.06)} & \multicolumn{2}{c}{1165 ($\pm$ 225) -- 865 ($\pm$115)} & \multicolumn{3}{c}{0.11 [$+$0.041, 37.2]} & \multicolumn{2}{c}{37.5 ($\pm$7.8)} & \multicolumn{2}{c}{10.2 ($-$19.6)} & \multicolumn{2}{c}{2.3 ($\pm$5.7)}\\
    \hline
    \multicolumn{14}{c}{Slow CMEs} \\
    \hline
    2008 Dec 12 & \multicolumn{2}{c}{0.13 ($\pm$0.006) -- 0.23($\pm$0.01)} & \multicolumn{2}{c}{235 ($\pm$ 40) -- 376 ($\pm$50)} & \multicolumn{3}{c}{0.05 [$+$0.008, 16]} & \multicolumn{2}{c}{18.7 ($\pm$3.6)} & \multicolumn{2}{c}{5 ($-$6)} & \multicolumn{2}{c}{1.3 ($\pm$1.1)} \\
    
    2010 Apr 03 & \multicolumn{2}{c}{0.26 ($\pm$0.04) -- 0.37 ($\pm$0.05)} & \multicolumn{2}{c}{461 ($\pm$40) -- 858 ($\pm$50)} & \multicolumn{3}{c}{0.22 [$+$0.057, 26]} & \multicolumn{2}{c}{27 ($\pm$4.3)} & \multicolumn{2}{c}{16.6 ($-$7.8)} & \multicolumn{2}{c}{14.2 ($\pm$4.5)}\\
    
    2010 Apr 08 & \multicolumn{2}{c}{0.14 ($\pm$0.02) -- 0.28 ($\pm$0.04)} & \multicolumn{2}{c}{290 ($\pm$ 23) -- 437 ($\pm$30)} & \multicolumn{3}{c}{0.068 [$+$0.002, 3]} & \multicolumn{2}{c}{21.2 ($\pm$4)} & \multicolumn{2}{c}{2.78 ($-$0.4)} & \multicolumn{2}{c}{2.3 ($\pm$0.3)} \\
    
    2011 Sep 13 & \multicolumn{2}{c}{0.34 ($\pm$0.05) -- 0.43 ($\pm$0.06)} & \multicolumn{2}{c}{216 ($\pm$15) -- 519 ($\pm$35)} & \multicolumn{3}{c}{0.1 [$+$0.027, 27]} & \multicolumn{2}{c}{30 ($\pm$5)} & \multicolumn{2}{c}{1.2 ($-$8.3)} & \multicolumn{2}{c}{2.8 ($\pm$3.5)}\\
    \hline 
    \end{tabular}
    \caption{Range (from the initial to the last tracked coronal height) of GCS-derived CME aspect ratio ($\kappa$), leading-edge speed ($V_{\rm LE}$), and their associated uncertainties/errors for the selected fast (top panel) and slow (bottom panel) CMEs. Uncertainties in $\kappa$ are based on the percentage errors derived from Table~\ref{tab:GCS_para} (see Section~\ref{sec:errorsgcs}), while uncertainties in $V_{\rm LE}$ are estimated from the corresponding fractional errors in the fitted heights. The table also lists the in situ measured $\kappa$ at 1 AU and associated uncertainties, obtained by correcting the measured radial size of MC for the spacecraft angular offset from the CME/MC propagation direction (Section~\ref{sec:erraspectmc}). The radial expansion-to-leading-edge speed ratio ($V_{\rm exp}/V_{\rm LE}$) is reported both at the final coronal height and at 1 AU. The speed ratio in COR is derived from GCS modeling-based speeds, with uncertainties propagated from speed errors. The in situ-measured speed ratio at 1 AU has corrections for spacecraft offset (Section~\ref{sec:errspdratio}). The speed ratio is also estimated using Equation~\ref{equ:vexp/vle}, with uncertainties derived from Equation~\ref{equ:vexp/vle_error}.}
    \label{tab:tab_2}
\end{table*}

\subsection{Uncertainties in GCS-Fitted Parameters}
\label{sec:errorsgcs}

The GCS technique is a forward-modeling approach in which a synthetic wireframe flux-rope geometry is visually matched to white-light CME observations. Consequently, uncertainties arise from user-dependent fitting choices and observer experience. \citet{Thernisien2009} noted that several GCS parameters are partially coupled, an intrinsic limitation of the fitting procedure. This parameter degeneracy is substantially reduced when two or three viewpoints with favorable spacecraft separation are used simultaneously \citep{Temmer2021,Verbeke2023,Nikou2025}. Additional uncertainties can result from low CME brightness, limited image cadence, background-subtraction quality, and the use of base- or running-difference images \citep{Thernisien2009,Bosman2012}. Another limitation of the GCS model is its simplified croissant-like flux-rope geometry for CMEs, introducing systematic uncertainties in the fitted parameters \citep{Vourlidas2013,Nikou2025}.

All of our selected events have been examined previously from different perspectives, and the reported 3D parameters show noticeable scatter among studies. For example, the selected fast CME of 2011 Jun 02 was analyzed by \citet{Zhuang2017,Temmer2021}, 2012 Jul 12 by \citet{Shen2014,Khuntia2024}, 2013 Apr 11 by \citet{Vemareddy2015,Sachdeva2017}, and 2013 Sep 29 by \citet{Kay2018,Temmer2021}. Similarly, the selected slow CMEs of 2008 Dec 12 were studied by \citet{Byrne2010,Mishra2013}, 2010 Apr 03 by \citet{Isavnin2013,Mishra2014}, 2010 Apr 08 by \citet{Kay2016,Temmer2021}, and 2011 Sep 13 by \citet{Sachdeva2017,Temmer2021}. It is evident that although the CMEs selected in this study are well observed and can be reliably tracked throughout the coronagraphic field of view, uncertainties in the derived GCS parameters are unavoidable.

In light of the above discussion, determining the true uncertainties in GCS fitting is challenging because they vary from event to event and can also evolve with time as the CME propagates. Different approaches have therefore been adopted in the literature. \citet{Thernisien2009} performed a sensitivity analysis and reported mean uncertainties of about $\pm1.8^\circ$ in latitude, $\pm4.3^\circ$ in longitude, $\pm22^\circ$ in tilt angle, $^{+13^\circ}_{-7^\circ}$ in half-angle, $\pm0.48\,R_{\odot}$ in height, and $^{+0.07}_{-0.04}$ in aspect ratio. Another approach is to fit the same CME independently by multiple researchers, as previously done with heliospheric imagers data \citep{Barnard2017}. Using this approach, \citet{Pluta2019} reported uncertainties of $\pm5^\circ$ in latitude, $\pm5^\circ$ in longitude, $\pm30^\circ$ in tilt angle, $\pm10^\circ$ in half-angle, $\pm0.5\,R_{\odot}$ in height, and $\pm0.025$ in aspect ratio.

More recently, \citet{Verbeke2023} used synthetic CME scenarios with known ``true'' parameters to directly quantify reconstruction errors for sufficiently evolved CMEs observed at heliocentric distances of $\sim$5--10 $R_{\odot}$. In their study, they reported representative minimum uncertainties for different spacecraft configurations: $6^{\circ\,+2^\circ}_{\,-3^\circ}$ in latitude, $11^{\circ\,+18^\circ}_{\,-6^\circ}$ in longitude, $25^{\circ\,+8^\circ}_{\,-7^\circ}$ in tilt angle, $10^{\circ\,+12^\circ}_{\,-6^\circ}$ in half-angular width, $0.6^{+1.2}_{-0.4}\,R_{\odot}$ in height, and $0.1^{+0.03}_{-0.02}$ in aspect ratio.

Despite these limitations, the GCS model remains one of the most widely used techniques for deriving CME 3D geometry and kinematics within the coronagraphic field of view when adequate multi-viewpoint observations are available, particularly for well-observed events such as those selected here. Based on repeated fits of the selected CMEs and the spread in our derived parameters, we find that our uncertainties remain within the ranges reported in previous studies \citep{Thernisien2009,Pluta2019,Verbeke2023}. In general, we observe that the propagation longitude, latitude, and height are among the best-constrained parameters, while the tilt angle typically carries the largest uncertainties. Among the size and shape parameters, the half-angle is typically more uncertain than the aspect ratio \citep{Sachdeva2017,Verbeke2023}.

All eight selected CMEs and their independent GCS reconstructions are available in the LLAMACoRe (Living List of Attributes Measured in Any Coronal Reconstruction; \citet{Kay2024}) super-database, which compiles measurements from several different sources of community CME catalogs, such as DONKI (Database Of Notifications, Knowledge, Information; \citet{Maddox2014}), the AFFECTS GCS catalog \citep{Bosman2012}. In the LLAMACoRe catalog, the selected fast CMEs of 2011 Jun 02, 2012 Jul 12, 2013 Apr 11, and 2013 Sep 29 have GCS reconstructions compiled from 10, 9, 7, and 8 independent published sources, respectively. Similarly, the selected slow CMEs of 2008 Dec 12, 2010 Apr 03, 2010 Apr 08, and 2011 Sep 13 have measurements compiled from 11, 14, 8, and 10 independent published sources, respectively. These independent reconstructions are used to derive a best-constrained estimate for each parameter, defined as the median value of the corresponding GCS-fitted parameter across all available sources. The associated uncertainty is represented by the median absolute deviation (MAD), calculated as the median of the absolute deviations from the median value.

We therefore compare our independently derived GCS parameters with these catalog median values and use the corresponding MADs as a practical estimate of observer-dependent uncertainties in our study. The reported MAD values are assumed to correspond to the final tracked stage of each event, as the LLAMACoRe catalog provides GCS parameters at the outermost distance/time when multiple reconstructions are available, i.e., when most of the coronal evolution of the CME has occurred \citep{Kay2024}. The uncertainties in different CME parameters are shown in Table~\ref{tab:GCS_para}. For GCS fitted parameters at our final tracked height, the reported MAD values are adopted as representative uncertainties for the corresponding fitted quantities. For longitude, latitude, and tilt angle, the corresponding MAD values (different for each event) are used as practical estimates of the uncertainties at all measured heights.

We estimate the uncertainty in the aspect ratio to be $\sim$5\% for the fast CME of 2012 Jul 12, $\sim$10\% for the 2013 Sep 29 CME, and $\sim$15\% for the 2011 Jun 02 and 2013 Apr 11 CMEs. For the slow CMEs, the uncertainty in the aspect ratio is estimated to be $\sim$15\% for all CMEs, except for the CME of 2008 Dec 12 for which it is $\sim$5\%. These fractional uncertainties are also listed in Table~\ref{tab:GCS_para}. The estimated uncertainties are assigned as error bars to the height--time measurements and subsequently propagated into the derived CME size and expansion speed estimates.

For the aspect ratio, we estimate the fractional uncertainty as $(\mathrm{MAD})_{\kappa}/\kappa_{f}$, which serves as a representative measure of the relative uncertainty/error in the fitted aspect ratio. This estimate is based on the catalog-listed MAD for $\kappa$ and our derived value of $\kappa$ at the final tracked height. We assume that the fractional uncertainty remains approximately constant during CME propagation, such that $\kappa$ at larger heights has proportionally larger absolute uncertainties. A similar approach is also used for the fitted angular width ($\alpha$) in our study.

The catalog does not explicitly report MAD (error) for the GCS-fitted heights, but it does report MAD for the CME speeds. We therefore use the fractional speed uncertainty, based on catalog values, as a representative relative uncertainty in the fitted heights. We assume that the relative uncertainty remains approximately constant for all data points. Based on this, we adopt a height uncertainty of approximately 10\% of the fitted height for the fast CMEs, except for the CME of 2013 Apr 11, for which the uncertainty is taken as $\sim$20\%. Similarly, for the slow CMEs, the fractional height uncertainty is assumed to be 5\%, except for the 2008 Dec 12 event, for which it is $\sim$10\%. These values are comparable to the representative height uncertainties of about 5--10\% reported in earlier CME studies \citep{Khuntia2024,Agarwal2024}.

We note that our GCS fitted values are generally consistent with the LLAMACoRe catalog best-constrained values within the corresponding MAD ranges \citep{Kay2024} and within uncertainties quoted in earlier studies \citep{Thernisien2009,Pluta2019,Verbeke2023}, providing additional confidence in the reliability of our reconstructions. In the present work, MAD-based uncertainties are shown as error bars in all relevant figures and are explicitly considered when interpreting the results. Accordingly, all trends and inferences for the selected fast and slow CMEs are presented in the following sections, with the associated parameter uncertainties taken into account.

\subsection{Aspect Ratio of MCs at 1 AU from In Situ Observations}
{\label{sec:aspectmc1au}}

The aspect ratio at 1 AU is determined using its relation with the MC radius and the heliocentric distance of the CME LE (Equation~\ref{equ:1}). The MC radius at the time of the LE arrival at 1 AU is estimated from in situ measurements using high-resolution (1-minute) OMNI data obtained from NASA CDAWeb (\url{https://cdaweb.gsfc.nasa.gov/}). Following \citet{Gopalswamy2015a,Agarwal2025}, the radial size of the MC is derived by integrating the bulk speed over the MC passage interval, which provides a reliable estimate under the assumption of a near-central (nose) encounter between the MC and the spacecraft.

Specifically, the MC radius ($R$) is calculated by integrating the speed–time profile over the duration of the MC using the trapezoidal rule,
\begin{equation} \label{equ:radialsize}
   R = \int_{t_0}^{t_n} V \, dt \approx \sum_{i = 0}^{n-1} \left(\frac{V_i + V_{i+1}}{2}\right) (t_{i+1} - t_i),
\end{equation}
where $V_i$ is the in situ measured speed at time $t_i$ ($i = 0, 1, 2, \ldots, n-1$) during the MC passage. The resulting radius is then used to estimate the aspect ratio at 1 AU for each CME.

\begin{figure*}
    \centering
    \includegraphics[scale= 0.37,trim={0cm 0cm 0cm 0cm},clip]{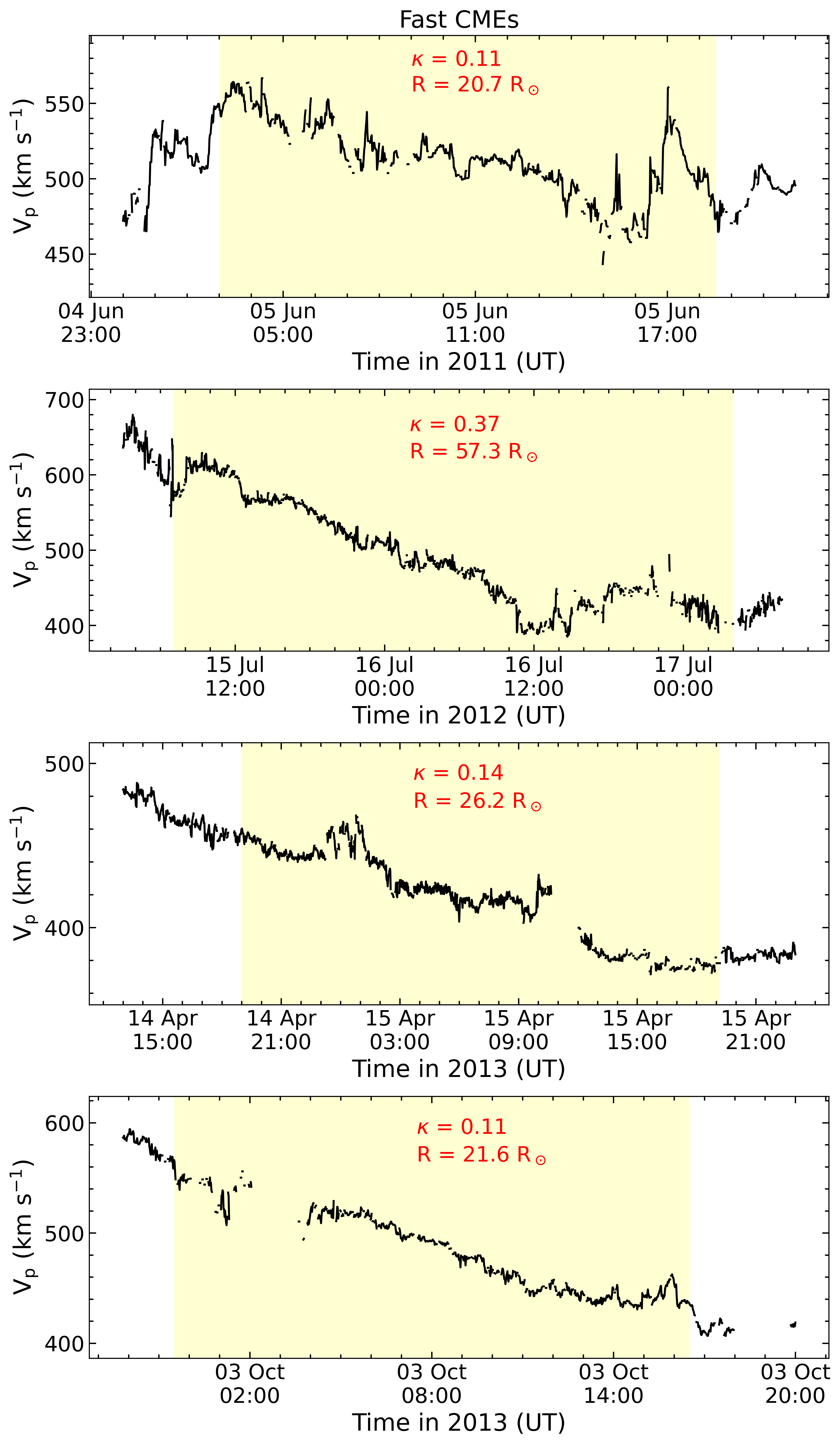}
   \includegraphics[scale=0.37,trim={0cm 0cm 0cm 0cm},clip]{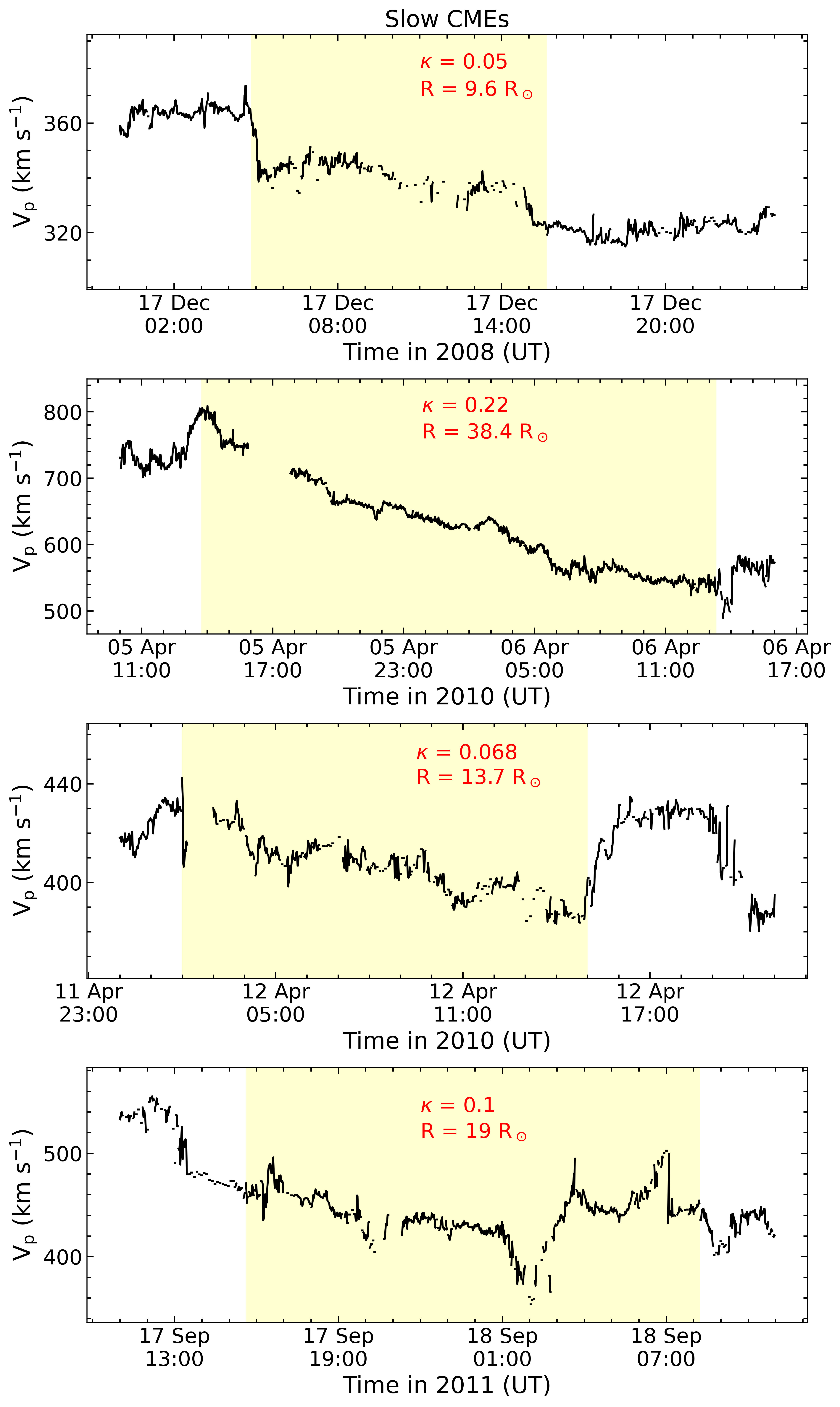}
    \caption{In situ measured proton speed ($V_p$) profiles of the selected fast (left column) and slow (right column) CMEs from OMNI data. The shaded regions mark the identified magnetic cloud (MC) intervals. The corresponding measured aspect ratio ($\kappa$) and MC radius ($R$) at 1 AU are annotated in each panel.}
    \label{fig:speed-time}
\end{figure*}

Figure~\ref{fig:speed-time} depicts the speed-time measurements of the fast and slow CMEs in the left and right panels, respectively, with the MC duration highlighted in the transparent yellow-shaded region. To ensure accurate boundary selection, we strictly considered regions with magnetic field rotation and plasma beta (ratio of the thermal pressure to the magnetic pressure) less than unity. On comparing the boundaries of identified MC with the boundaries of ICME listed in Richardson and Cane's ICME catalog (\url{https://izw1.caltech.edu/ACE/ASC/DATA/level3/icmetable2.htm}), we find a deviation of at most $\pm$2 hours. We note that such a deviation can arise because ICME boundaries in the catalog are derived at L1, while in OMNI data, which we rely on, make time-shifts L1 measurements to the Earth’s bow shock \citep{King2005}. Further, the ICME boundaries in the catalog are based on solar wind composition and charge-state measurements from the ACE spacecraft \citep{Richardson2004}, whereas our MC boundaries are primarily based on magnetic field measurements \citep{Lepping1990}.

For the MCs associated with the selected CMEs, the in situ measured radius at 1 AU is listed in the third column of Table~\ref{tab:del_corr_R_vle}. Each panel of Figure~\ref{fig:speed-time} lists (with red) the estimated value of $R$ and $\kappa$ at 1 AU for each fast and slow CME. The estimated value of $R$ of each MC at 1 AU is used in Equation~\ref{equ:1} to estimate the $\kappa$. The fourth column of Table~\ref{tab:tab_2} lists the in situ measured $\kappa$ at 1 AU and associated uncertainties (correction needed) for each event. As discussed in Section~\ref{sec:erraspectmc}, these uncertainties are based on the fact that the spacecraft trajectory deviating from the MC center underestimates the $R$ and $\kappa$ of the MCs. The values in the square brackets of the fourth column in Table~\ref{tab:tab_2} indicate the correction required, expressed as both the absolute increase in $\kappa$ and the corresponding percentage increase relative to the measured value, to account for the spacecraft’s offset from the MC center.

Based on the situ measured $\kappa$ at 1 AU and the required corrections into it as listed in Table~\ref{tab:tab_2}, we note that percentage corrections in $\kappa$ for the fast CMEs of 2011 Jun 02, 2012 Jul 12, 2013 Apr 11, and 2013 Sep 29 are $\sim22\%$, $\sim5\%$, $\sim11\%$, and $\sim37\%$, respectively. Similarly, the percentage corrections in $\kappa$ for the slow CMEs of 2008 Dec 12, 2010 Apr 03, 2010 Apr 08, and 2011 Sep 13 are $\sim16\%$, $\sim26\%$, $\sim3\%$, and $\sim27\%$, respectively. These results indicate that uncertainties in the aspect ratio arising from the spacecraft crossing geometry can, in some cases, exceed the typical uncertainties associated with the GCS-derived aspect ratio estimates (Table~\ref{tab:GCS_para}) at coronal heights.

Based on the in situ measured parameters at 1 AU, the MCs associated with fast CMEs are, on average, larger and exhibit higher aspect ratios than those associated with slow CMEs. The mean measured radii for the slow and fast events are $\sim20.2(+3.8)$ and $\sim31.5 (+3.5) \,R_{\odot}$, respectively, while the corresponding mean aspect ratios are $\sim0.11(+0.02)$ and $\sim0.18(+0.02)$. The uncertainties/corrections are based on the description in Section~\ref{sec:erraspectmc}, accounting for the spacecraft angular offset from the CME propagation direction. The average radial size of fast and slow MCs is smaller than that reported in earlier studies \citep{Lepping1990,Liu2005,Mishra2021a,Zhuang2023}. This difference may arise from the limited sample size considered in our study, the event-selection criteria favoring well-observed Sun--Earth CMEs, and possible solar-cycle dependence. In the following section, we discuss the possibility that these in situ-measured radial sizes and aspect ratios may require correction, as they are based on the measured MC size, which can be underestimated when the spacecraft trajectory deviates from the MC center.

\begin{table}
    \centering
    \setlength{\tabcolsep}{3.5pt}
    \footnotesize
    \begin{tabular}{cccc}
    \hline
    \multirow{3}{*}{Date} & \multirow{3}{*}{$\Delta~(^\circ)$} & {In situ measured} & In situ measured\\
     & & R $(R_\odot)$ at 1 AU & $V_{LE}~(km~s^{-1})$ at 1 AU \\
     & & [Correction (abs.,\% )] & [Correction (abs.,\% )]\\
    \hline
    \multicolumn{4}{c}{Fast CMEs} \\
    \hline
    2011 Jun 02 & 18 & 20.7 [+4.4, 21.3] & 546 [+27, 4.9] \\
    2012 Jul 12 & 15 & 57.3 [+2.6, 4.5] & 573 [+17, 3] \\
    2013 Apr 11 & 20 & 26.2 [+2.4, 9.1] & 456 [+27, 5.9] \\
    2013 Sep 29 & 38 & 21.6 [+6.3, 29.1] & 549 [+133, 24]\\
    \hline
    \multicolumn{4}{c}{Slow CMEs} \\
    \hline
    2008 Dec 12 & 18 & 9.6 [+2.1, 21.9] & 358 [+18, 5]\\
    2010 Apr 03 & 24 & 38.4 [+7.8, 20.3] & 801 [+63, 7.9] \\
    2010 Apr 08 & 7 & 13.7 [+0.1, 0.73] & 414 [+0.8, 0.19]\\
    2011 Sep 13 & 28 & 19 [+5, 26.3] & 460 [+60, 13]\\
    \hline 
    \end{tabular}
    \caption{The table lists the spacecraft angular offset ($\Delta$) from the CME/MC propagation direction, in situ measured MC radius at 1 AU, along with corresponding uncertainties (see Section~\ref{sec:erraspectmc}). It also includes the in situ measured LE speed at 1 AU, along with corresponding uncertainties (see Section~\ref{sec:errspdratio}).}
    \label{tab:del_corr_R_vle}
\end{table}

\subsection{Uncertainties and Corrections in Estimated Aspect Ratio of MCs at 1 AU}{\label{sec:erraspectmc}}

The MC radius and aspect ratio estimated above from the in situ speed--time profiles are subject to geometric selection effect arising from the spacecraft trajectory relative to the MC center \citep{Zhang2013}. Such limitations of single-spacecraft measurements are well recognized, as the inferred MC size depends on the spacecraft’s impact parameter \citep{Lepping1990,Lynch2003,Song2020}.  The observed passage duration of MC at the spacecraft typically represents a chord through the MC cross-section rather than the full central diameter. As a result, the radius ($R$) derived from in situ observations may be underestimated when the spacecraft does not pass through the MC center \citep{Demoulin2013}. Since the aspect ratio derived in this work depends directly on the MC radius estimated at 1 AU, this geometric bias must be accounted for when comparing the aspect ratios of selected CMEs between near the Sun and near the Earth.

To quantify this effect, we adopt an idealized geometrical model in which the MC cross-section at the spacecraft distance is represented by a circle of true radius $R_{\rm t}$, whose center is located at a heliocentric distance $D$. Although the selected CMEs are Earth-directed and sampled as MCs by an L1 spacecraft, their propagation directions inferred from GCS reconstructions are not necessarily aligned with the Sun--Earth line. We assume that the MC at 1 AU has the same direction as inferred from the GCS model at the final tracked coronal height. We estimate the angular separation ($\Delta$) between the MC propagation direction ($\theta_{\rm MC},\phi_{\rm MC}$) and the spacecraft location ($\theta_{\rm sc},\phi_{\rm sc}$) through,
$\cos\Delta =
\sin\theta_{\rm MC}\sin\theta_{\rm sc}
+
\cos\theta_{\rm MC}\cos\theta_{\rm sc}\cos(\phi_{\rm MC}-\phi_{\rm sc})$

The angular separation ($\Delta$) corresponds to a perpendicular miss distance (impact parameter) between the spacecraft trajectory and the MC center given by $b$ = $D\sin\Delta$, where $b$ is the shortest distance between the spacecraft path and the MC center. If the MC has an angular half-width $\lambda$, then its true radius ($R_{\rm t}$) at distance $D$ is $R_{\rm t}=D\sin\lambda$. A spacecraft encounter occurs only when the trajectory intersects the MC cross-section, i.e., when $b \leq R_{\rm t}$, or equivalently when $\Delta \leq \lambda$. Thus, $\Delta=0$ represents a central hit and $\Delta \rightarrow \lambda$ approaches a grazing encounter near the boundary of MC. Therefore, the radius ($R$) measured from the in situ duration corresponds to the half-chord length rather than the true radius $R_{\rm t}$. From simple circle geometry, the relation between the measured and true radius would be as, $R$ = $\sqrt{R_{\rm t}^{2}-b^{2}} = \sqrt{R_{\rm t}^{2}-D^{2}\sin^{2}\Delta}$. Thus, $R=R_{\rm t}$ for a central encounter ($\Delta=0$), whereas $R<R_{\rm t}$ for any non-zero offset. The underestimated measured radius ($R$) can be written as follows:

\begin{equation}
{R = {R_{\rm t}} \left(\sqrt{1-\left(\frac{\sin\Delta}{\sin\lambda}\right)^{2}}\right)}
\end{equation}

Using the MC's angular separation ($\Delta$) relative to the in situ spacecraft (listed in Table~\ref{tab:del_corr_R_vle}), we calculated the true MC radius ($R_{\rm t}$) and aspect ratio. For the fast CMEs of 2011 Jun 02, 2012 Jul 12, 2013 Apr 11, and 2013 Sep 29, the required corrections to the measured radius correspond to percentage increases of approximately $21\%$, $4.7\%$, $9\%$, and $29.6\%$, respectively, as listed in Table~\ref{tab:del_corr_R_vle}. Likewise, for the slow CMEs of 2008 Dec 12, 2010 Apr 03, 2010 Apr 08, and 2011 Sep 13, the required corrections correspond to percentage increases of approximately $22\%$, $20\%$, $2\%$, and $26.3\%$, respectively. Based on these corrections, the estimated correction magnitudes and their corresponding percentages for the aspect ratio at 1 AU are described in Section~\ref{sec:aspectmc1au} and listed in Table~\ref{tab:tab_2}. The results enable a direct comparison of the systematic bias introduced by flank encounters in different CMEs. The mean corrected true radius for the fast and slow MCs are $\sim35$ and $\sim24~R_{\odot}$, respectively, while the corresponding mean aspect ratios increase to $\sim0.20$ and $\sim0.13$. These results suggest that fast CMEs tend to evolve into larger, more radially extended structures by 1 AU. The uncertainties discussed above from the geometric selection effect are further considered in our analysis of the evolution of aspect-ratio with heliocentric distance.

\subsection{Three-Phase Progression of Aspect Ratio of CMEs}
\label{sec:expaspect}

\begin{figure*}
    \centering
    \includegraphics[scale= 0.35,trim={0cm 0cm 0cm 0cm},clip]{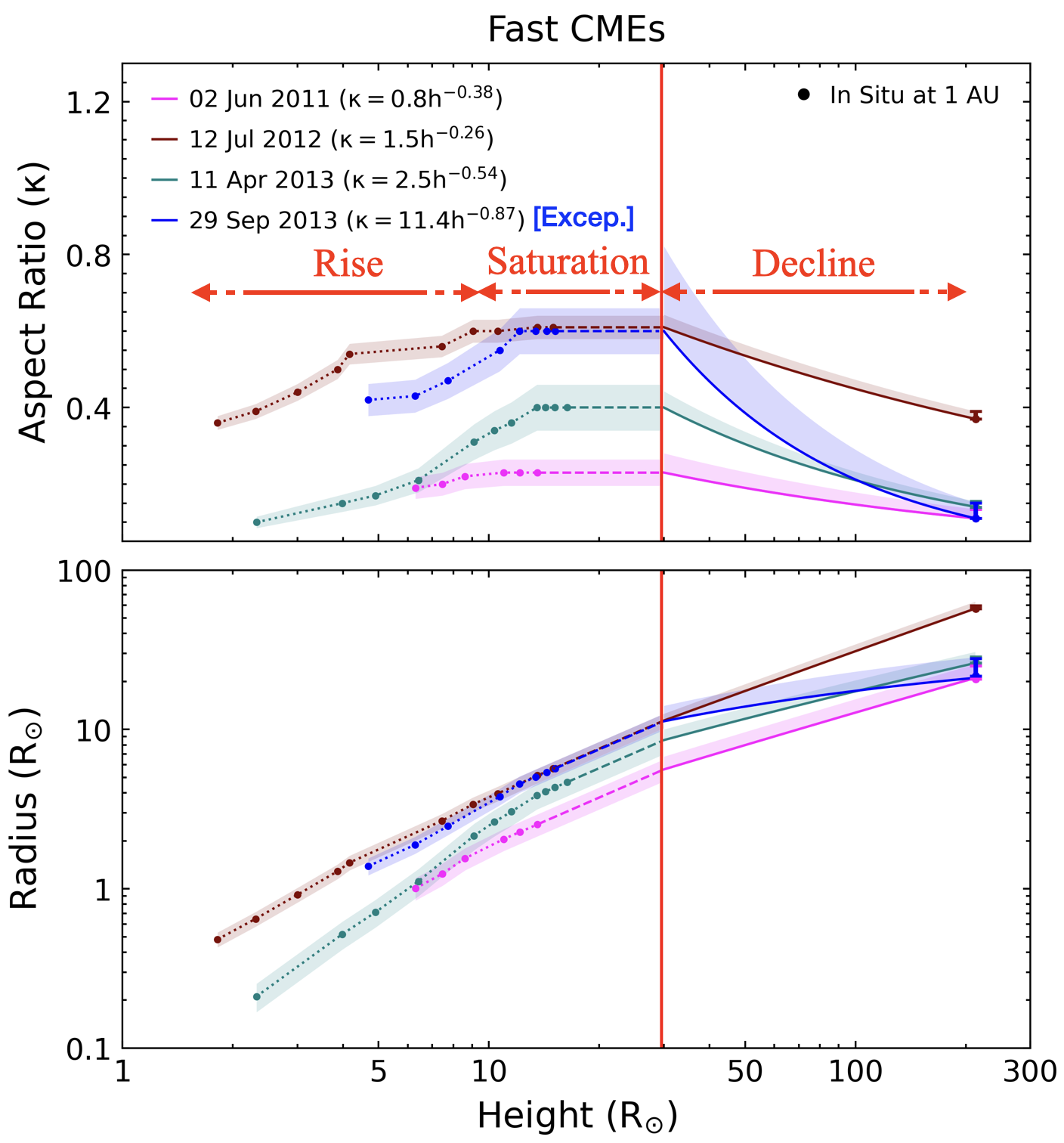}
   \includegraphics[scale=0.35,trim={0cm 0cm 0cm 0cm},clip]{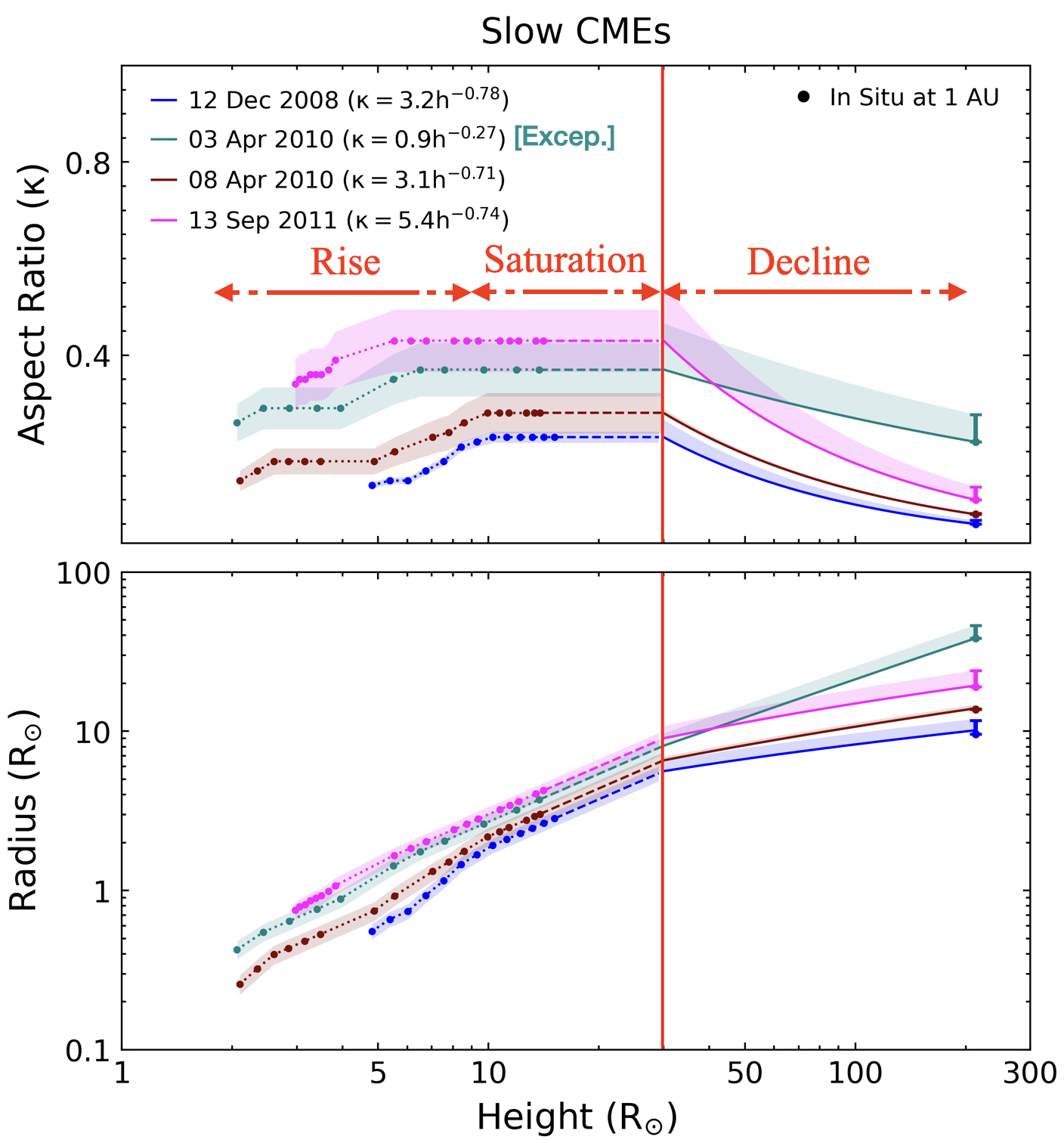}
   \caption{Evolution of the CME aspect ratio ($\kappa$; top panels) and radius ($R$; bottom panels) as a function of heliocentric distance for the selected fast (left panels) and slow (right panels) CMEs. The larger dot symbols over the dotted lines represent data points derived from GCS modeling, with different colors indicating individual events. The dashed lines show the extrapolated $\kappa$ based on GCS measurements, while the solid curves represent power-law fits constrained between the values at $\sim30\,R_{\odot}$ (marked with red vertical line) and the in situ measurements at 1 AU. The percentage uncertainties in the GCS-derived $\kappa$ (Section~\ref{sec:errorsgcs}) are indicated by shaded regions and are assumed constant up to $\sim30\,R_{\odot}$. The percentage uncertainties in the in situ $\kappa$ at 1 AU (Section~\ref{sec:erraspectmc}) are shown as error bars and are also applied to the fitted power-law estimates. One CME in each category is marked as [Excep.] to denote its exceptional (outlier) behavior, exhibiting an unusually different power-law index during the decline phase compared to the remaining events in the respective group.}
    \label{fig:aspect_ratio_evolution}
\end{figure*}

We examine the evolution of the CME aspect ratio from near the Sun to 1 AU. Figure~\ref{fig:aspect_ratio_evolution} shows the variation of aspect ratio (top panels) and radius (bottom panels) with CME LE height for the selected fast (left) and slow (right) CMEs. The dots over the dotted line represent each data point of the aspect ratio estimated from the GCS model. The shaded regions indicate the associated percentage uncertainties as discussed in Section~\ref{sec:errorsgcs} and listed in Table~\ref{tab:GCS_para}. It shows that the aspect ratio initially increases, then becomes constant towards the last tracked/observed height of approximately 15 $R_\odot$ for all CMEs. The initial increase in the aspect ratio is more pronounced for fast CMEs, whereas the approach to the constant phase is clearer for slow CMEs. We note that the aspect ratio (with uncertainties as discussed in Section~\ref{sec:aspectmc1au} at 1 AU from in situ observations (shown with filled circles) for each CME is lower than its value from the GCS model at the last tracked height.

It is possible that the aspect ratio ($\kappa$) of the CMEs first increases for some height and then remains constant, followed by a decrease in the interplanetary medium as suggested in \citet{Agarwal2024}. We follow the study of \citet{Agarwal2024} and assume that the value $\kappa$ beyond our tracked height remains constant up to a certain distance, such as the height of 30 $R_\odot$, as shown by the dashed line in the figure, before decreasing in the interplanetary medium. We use a power law fit, shown by the solid line in the figure, to infer the interplanetary evolution of $\kappa$ beyond 30 $R_\odot$ (marked with red vertical solid line) to 1 AU. The percentage uncertainties (shown with a shaded region) in the constant phase are taken as equal to those in the observed rise phase. The uncertainties (shown as shaded region) in the $\kappa$ inferred from the fitted power law are taken to be the same as the percentage uncertainties (corrections shown as error bars) in the $\kappa$ at L1 measured in situ.

From Figure~\ref{fig:aspect_ratio_evolution}, the evolution of the CME aspect ratio ($\kappa$) can be described in three phases: an initial \textit{rise}, followed by a \textit{saturation}, and a subsequent \textit{decline} toward 1 AU. The rise phase, observed in the low and middle corona ($\lesssim 10$--$15\,R_{\odot}$), is directly constrained by GCS-model measurements and indicates rapid radial expansion of CMEs. This increase in $\kappa$ is more pronounced for fast CMEs, reflecting stronger expansion dynamics. At intermediate heights ($\sim 15$--$30\,R_{\odot}$), $\kappa$ approaches the saturation phase, as inferred from the leveling-off behavior toward the end of the observed rise, suggesting a transition to a more balanced expansion regime. Beyond this, the decline phase is constructed by incorporating the in situ measured aspect ratio at 1 AU and its deviation from near-Sun estimates. The observed decrease in $\kappa$ toward 1 AU implies a reduction in radial expansion of CMEs during interplanetary propagation. This behavior indicates three distinct evolutionary phases of the aspect ratio. While both fast and slow CMEs follow this general trend, fast CMEs attain higher peak $\kappa$ values and exhibit larger variations, whereas slow CMEs show a more gradual rise, lower saturation levels, and a comparatively faster decline \citep{Cremades2020}.

Beyond $\sim 30\,R_{\odot}$, the decline in aspect ratio is characterized by power-law relations of the form $\kappa \propto h^{n}$. For fast CMEs, the fitted indices range from $n \sim -0.26$ to $-0.87$, with a mean value of $\sim -0.51$. Considering the uncertainties in the fitted power-law, the CME on 2013 Sep 29 is identified as an exceptional event among the fast CMEs because it exhibits a significantly different (steeper) power-law index during the decline phase compared to the remaining events in the group. Excluding this case, the mean index becomes $\sim -0.39$. It remains to be investigated whether the inferred rapid decrease for the exceptional CME is influenced by the relatively larger uncertainties in the estimated $\kappa$ at 1 AU, since the CME propagated about $30^\circ$ west of the Sun--Earth line and may also have experienced additional deflection. Other factors that can introduce uncertainties in the measured $\kappa$ values from remote and in situ observations are discussed in Section~\ref{sec:resdis}.

In contrast, slow CMEs exhibit fitted indices ranging from $n \sim -0.27$ to $-0.78$, with a mean value of $\sim -0.63$. Even if we account for the uncertainties (Table 2 and Section~\ref{sec:erraspectmc}) in the aspect ratio as fitted with a power-law, the CME on 2010 Apr 03 is identified as an exceptional event among the slow CMEs because it exhibits a significantly different (shallower) power-law index during the decline phase compared to the remaining events in the group. Excluding this, the mean index becomes $\sim -0.74$. Earlier studies have suggested that this CME underwent minimal deceleration in the interplanetary medium due to its interaction with the high-speed ambient solar wind \citep{Liu2011,Mishra2014,Agarwal2024}. Based on the mean values of the power-law indices, we infer that fast CMEs tend to exhibit comparatively shallower declines than slow CMEs, although one exceptional case is present in each group.

From the above analysis, the $\kappa$ is found to rise initially, approach a saturation, and subsequently decline with increasing CME height. To capture this behavior, we model the slightly smoothed values of $\kappa$ in Figure~\ref{fig:aspect_ratio_evolution} using two functions, each composed of three terms representing the rise, saturation, and decline phases. These functions (Equations~\ref{equ:func1} and \ref{equ:func2}) are fitted to the estimated aspect ratio profiles for both fast and slow CMEs from near the Sun to 1 AU.

\begin{align} \label{equ:func1}
    f_1(h,\kappa) = A \left(\frac{h}{a} \right)^n \left(1 - \frac{1}{1 + e^{-k_1 h}}\right) 
    + B \left(\frac{1}{1 + e^{-k_1 h}} \right) \nonumber\\ 
    \left(1 - \frac{1}{1 + e^{-k_2 h}}\right) 
    + C \left(\frac{h}{b} \right)^{-m} \frac{1}{1 + e^{-k_2 h}}
\end{align}

In the Equation~\ref{equ:func1}, the first term describes the initial increase in $\kappa$ as a power-law function of height, scaled by $A$ and exponent $n$, with $a$ providing a normalization length scale. Its contribution is smoothly reduced at larger heights by the logistic transition controlled by $k_1$. The second term, scaled by $B$, represents the intermediate saturation phase and is active primarily between the two transition heights defined by $k_1$ and $k_2$. The third term describes the decline in $\kappa$ at larger distances as a decreasing power law with exponent $m$, scaled by $C$ and normalized by $b$, and becomes dominant beyond the second transition controlled by $k_2$. The logistic functions, therefore, act as smooth switching functions that ensure a continuous transition between the three evolutionary phases.

\begin{equation} \label{equ:func2}
    f_2(h,\kappa) = A \left(\frac{h}{a} \right)^n e^{-\beta (\frac{h}{a})^{\gamma}} \left(1 + \frac{h}{b}\right)^{-\delta}
\end{equation}

Equation~\ref{equ:func2} represents the evolution of $\kappa(h)$ using a single continuous function that captures the rise, saturation, and decline phases. The initial increase in $\kappa$ at lower heights is governed by the power-law term $A (h/a)^n$, where $A$ is a scaling factor, $a$ is a normalization length scale, and $n$ controls the rate of growth. The exponential term $\exp[-\beta (h/a)^{\gamma}]$ at intermediate heights effectively produces a saturation-like behavior, where $\beta$ controls the strength of damping and $\gamma$ determines how rapidly the suppression sets in with height. The decline phase at larger distances is controlled by the term $(1 + h/b)^{-\delta}$, where $b$ sets the transition scale and $\delta$ controls the rate of decay. This function, without having explicit switching functions, can provide a smooth and continuous evolution of the aspect ratio as observed between near-Sun to 1 AU.

\begin{figure*}
    \centering
    \includegraphics[scale= 0.82,trim={0cm 0cm 0cm 0cm},clip]{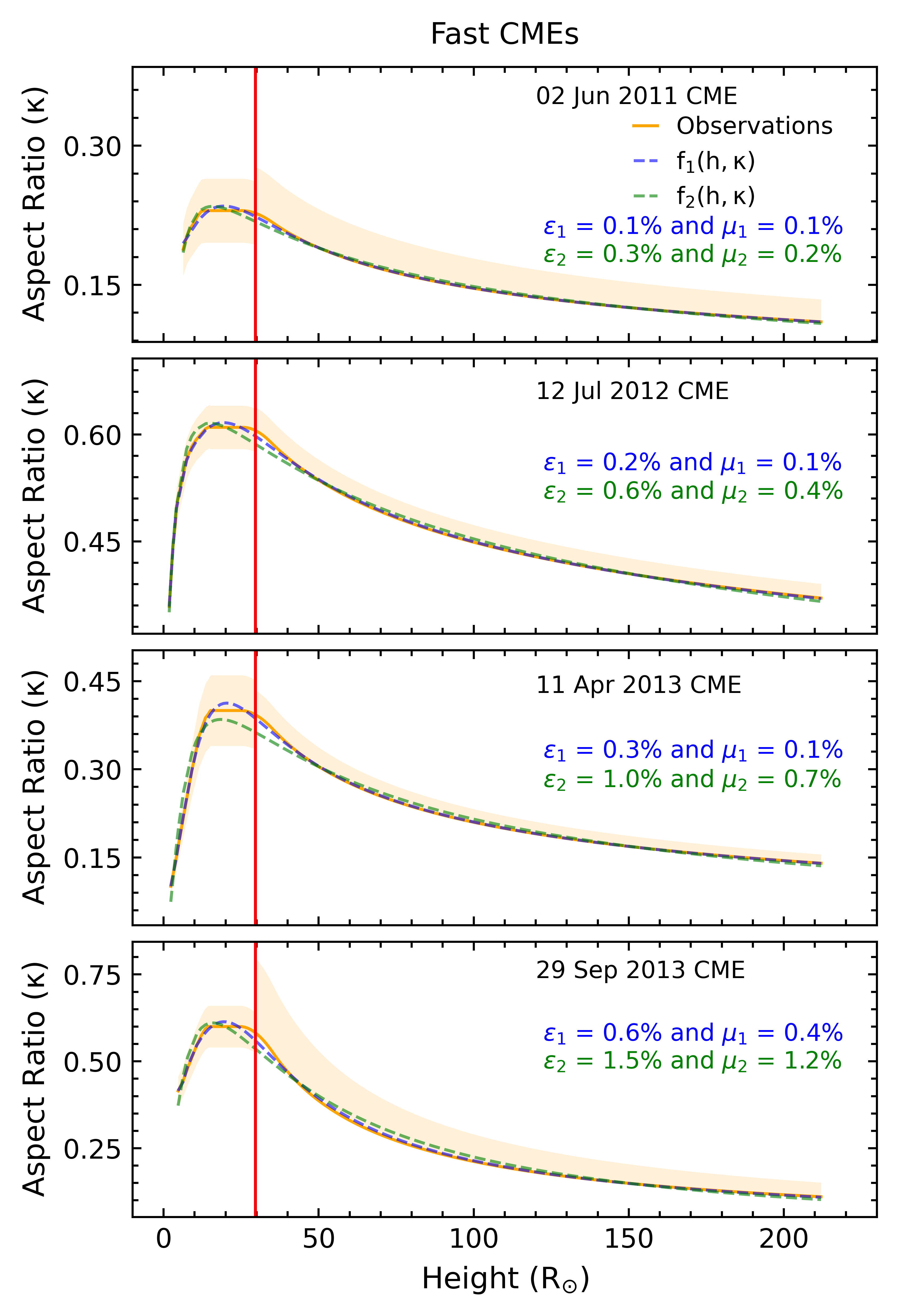}
   \includegraphics[scale=0.82,trim={0cm 0cm 0cm 0cm},clip]{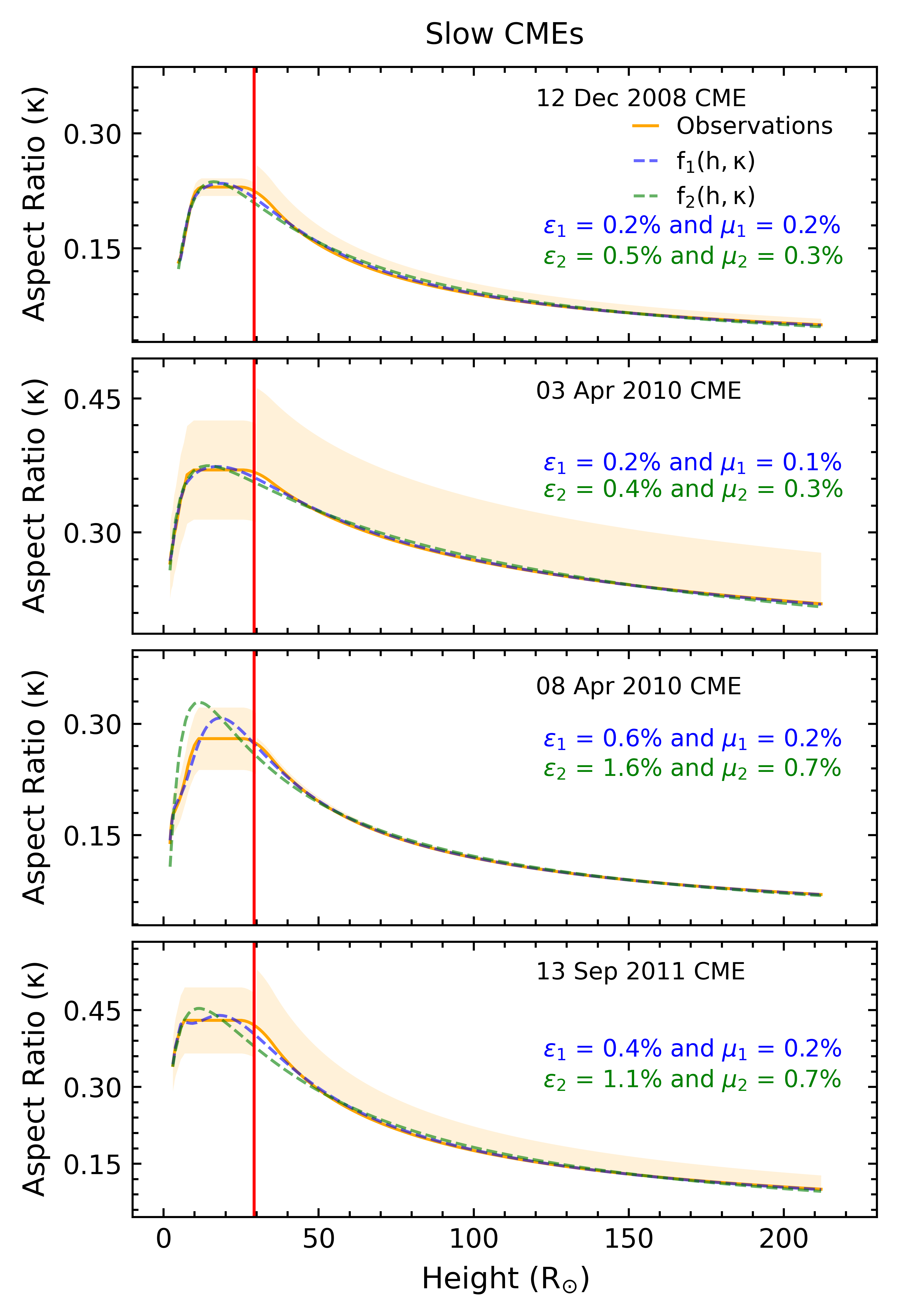}
   \caption{Model fits to the evolution of CME aspect ratio ($\kappa$) with heliocentric distance for fast (left panels) and slow (right panels) CMEs. The orange curve represents the observed profile, combining GCS-derived values in the corona and the inferred evolution in the interplanetary medium, with a shaded region indicating the same uncertainties as shown in Figure~\ref{fig:aspect_ratio_evolution}. The blue and green curves show fits using the proposed functions $f_1(h,\kappa)$ and $f_2(h,\kappa)$, respectively. The vertical red line marks the transition at $\sim30\,R_{\odot}$. The root mean square error ($\epsilon$) and mean absolute error ($\mu$) for each fit are indicated in the panels.}
    \label{fig:kap_funct_fast_slow}
\end{figure*}

Figure~\ref{fig:kap_funct_fast_slow} shows the modeled evolution of the CME aspect ratio ($\kappa$) as a function of heliocentric distance for fast and slow CMEs, displayed in the left and right panels, respectively. The orange curves represent the reference (based on observations) aspect ratio profiles, which are the slightly smoothed values of $\kappa$ in Figure~\ref{fig:aspect_ratio_evolution}. The fractional uncertainties shown with the shaded region are the same as described in Section~\ref{sec:expaspect} and shown in Figure~\ref{fig:aspect_ratio_evolution}.

The fits are obtained using a weighted least-squares approach, accounting for the uncertainties in the data points. The uncertainties in the data points are supplied as $\sigma$ and each point is weighted by the inverse of its variance ($w_i = 1/\sigma_i^2$). The blue and green curves correspond to fits obtained using the proposed functions $f_1(h,\kappa)$ (Equation~\ref{equ:func1}) and $f_2(h,\kappa)$ (Equation~\ref{equ:func2}), respectively, applied over the full range from near the Sun to 1 AU. The vertical red line at $\sim30\,R_{\odot}$ marks the transition between coronal and interplanetary propagation regimes. Both functions reproduce the overall rise–saturation–decline behavior of the aspect ratio. The quality of the fits is quantified using the root mean square error ($\epsilon$) and mean absolute error ($\mu$), which are shown in each panel. The close agreement between the fitted curves and the reference/observed profile confirms that the adopted parameterizations provide an adequate description of the aspect ratio evolution within the assumed uncertainties.

Overall, both functions provide good agreement with the observed profiles, with small values of $\epsilon$ and $\mu$ across all events. However, $f_1(h,\kappa)$ consistently yields slightly lower errors, indicating a marginally better representation of the aspect ratio evolution compared to $f_2(h,\kappa)$. This suggests that an explicit representation of distinct evolutionary phases, with controlled transitions between them, may more effectively capture the observed behavior than a fully continuous formulation.

We find that the evolution of the CME aspect ratio ($\kappa$) is governed by a sequence of regimes. In the \textit{rise phase} of $\kappa$, the expansion is dominated by rapid radial growth driven by strong internal magnetic pressure. This is followed by a \textit{saturation phase} of $\kappa$, characterized by a regulated expansion, in which the decline of internal pressure and the ambient solar wind pressure becomes comparable. Finally, in the \textit{decline phase} of $\kappa$, the internal overpressure relative to the ambient medium weakens, and the expansion is increasingly influenced by the surroundings, leading to a reduced radial expansion efficiency. This progression reflects a transition from magnetically dominated expansion to a regulated expansion regime, and ultimately to one controlled by the heliospheric environment.

Since the aspect ratio directly relates to the geometry of conical legs in croissant-shape CME flux rope, it is linked to the expansion of CME along both radial and lateral dimensions of CME propagation. The observed three-phase evolution of the GCS-derived aspect ratio suggests that CMEs do not undergo strict self-similar expansion from the Sun to 1 AU. Under self-similar expansion, the aspect ratio is expected to remain nearly constant. Instead, the observed rise, saturation, and decline in $\kappa$ indicate significant geometric evolution during propagation. The saturation phase may correspond to self-similarity, strictly for conical legs, while the rise and decline phases likely reflect significant changes in internal pressure forces and interactions with the ambient solar wind.

\vspace{-1mm}
\subsection{Evolution of the Expansion-to-Leading-Edge Speed Ratio of CMEs}
\label{spdratio}

To further investigate the expansion behavior of CMEs, we estimate the ratio of radial expansion speed to LE speed both near the Sun (at the final tracked height in COR) and at 1 AU. We estimate the ratio of the radial expansion speed to the LE speed at 1 AU using two methods. The first method is by using in situ speed-time measurements, and the conventional method for finding the expansion speed as $V_{exp} = \frac{V_L - V_T}{2}$, where $V_L$ and $V_T$ are the LE and trailing edge (TE) speed of the CME \citep{Owens2005,Gopalswamy2014}. However, this expansion speed is not instantaneous at the arrival time of LE on the in situ spacecraft \citep{Agarwal2024,Agarwal2026}. The second method to estimate the ratio of expansion speed to the LE speed ($V_{exp}/V_{LE}$) at 1 AU is by taking the derivative of Equation~\ref{equ:1} with respect to the height as follows:

\begin{align}\label{equ:vexp/vle}
    \frac{dR}{dh} & = \frac{\kappa}{1+\kappa} + h\left(\frac{1}{1+\kappa} - \frac{\kappa}{(1 + \kappa)^2}\right)\frac{d\kappa}{dh} \nonumber \\
   \frac{dR}{dt} & \frac{dt}{dh}  = \frac{\kappa}{1+\kappa} + \left(\frac{h}{(1+\kappa)^2}\right)\frac{d\kappa}{dh} \nonumber \\
   &\frac{V_{exp}}{V_{LE}} =\frac{\kappa}{1+\kappa} + \left(\frac{h}{(1+\kappa)^2}\right)\frac{d\kappa}{dh}
\end{align}

\begin{figure*}
    \centering
    \includegraphics[scale= 0.82,trim={0cm 0cm 0cm 0cm},clip]{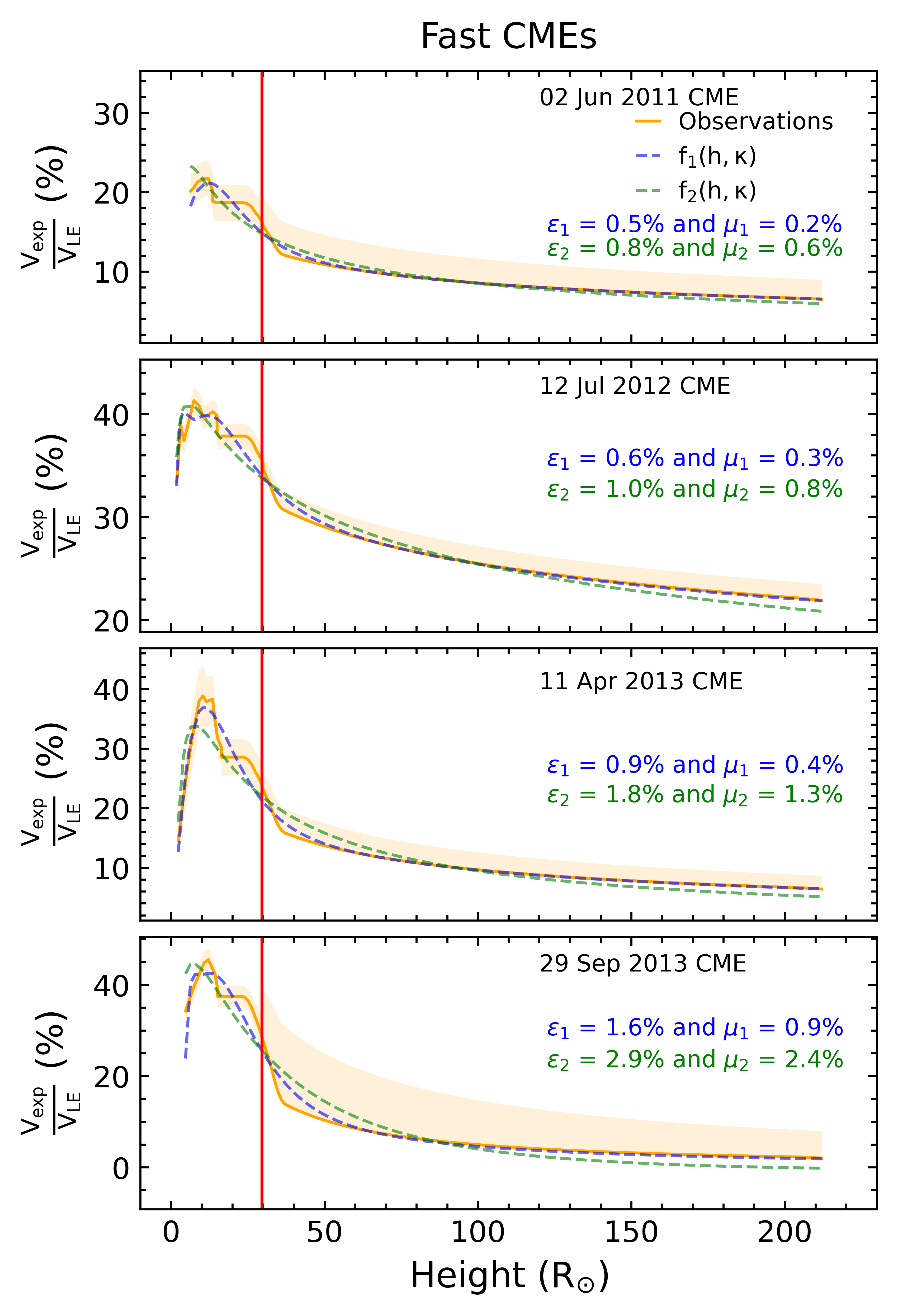}
   \includegraphics[scale=0.82,trim={0cm 0cm 0cm 0cm},clip]{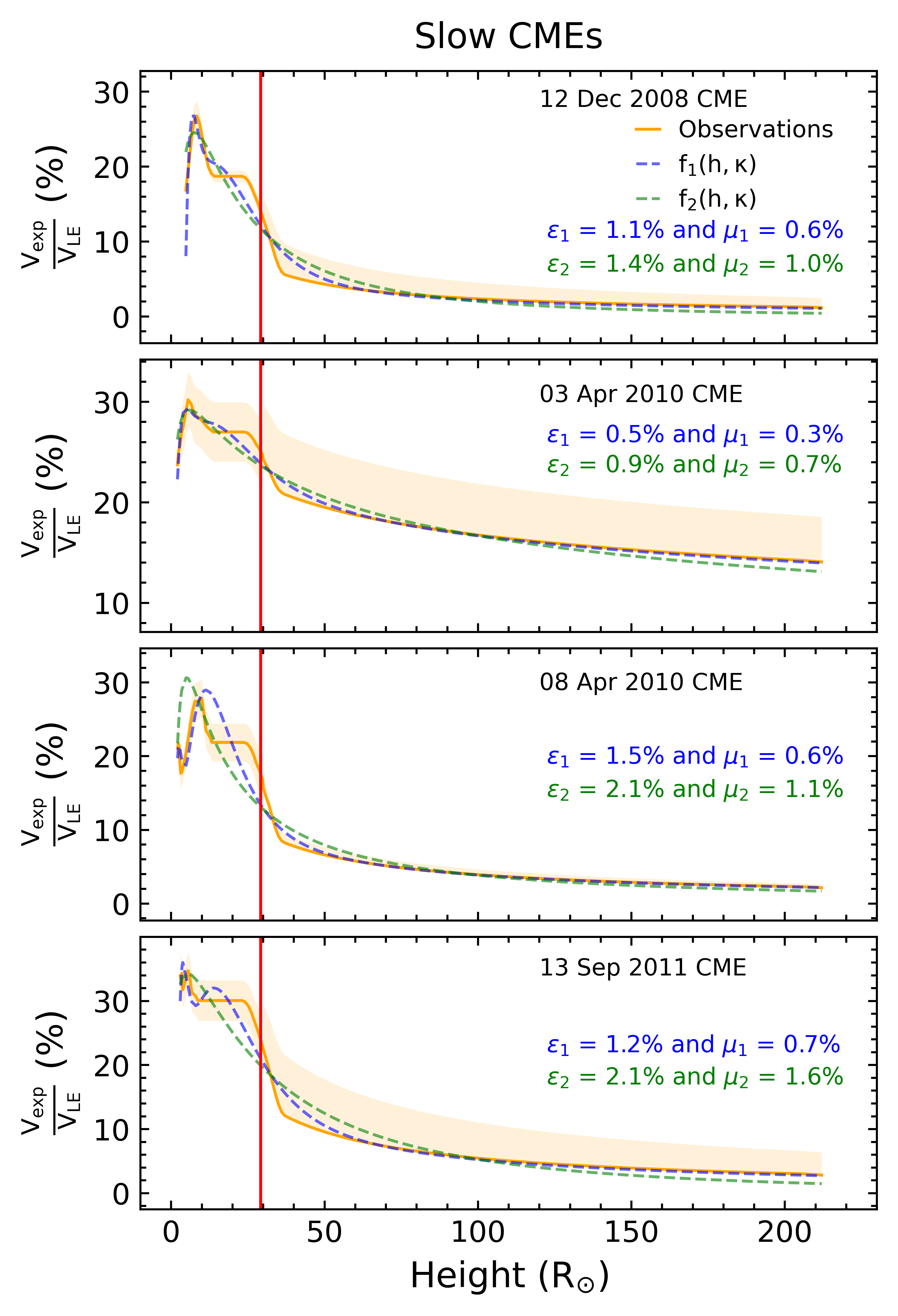}
   \caption{Evolution of the ratio of radial expansion speed to leading-edge speed ($V_{\rm exp}/V_{\rm LE}$) as a function of heliocentric distance for fast (left panels) and slow (right panels) CMEs. The orange curves represent the observed profiles, with shaded regions indicating uncertainties (using standard error propagation described in Section~\ref{sec:errspdratio}). The blue and green curves show fits obtained using the functions $f_1'(h,\kappa)$ and $f_2'(h,\kappa)$, respectively. The vertical red line at $\sim30\,R_{\odot}$ marks the transition from coronal to interplanetary propagation. The root mean square error ($\epsilon$) and mean absolute error ($\mu$) are indicated in each panel.}
    \label{fig:vexp_vle_funct}
\end{figure*}

Utilizing the derivative of slightly smoothed values of $\kappa$ in Figure~\ref{fig:aspect_ratio_evolution}, we estimate the value of $\frac{V_{exp}}{V_{LE}}$ from Equation~\ref{equ:vexp/vle} for each CME from near the Sun to 1 AU. Figure~\ref{fig:vexp_vle_funct} shows the estimated value of $V_{\rm exp}/V_{\rm LE}$ (with solid orange curve) for fast (left panels) and slow (right panels) CMEs. The orange curve represents the reference/observed profile derived from observations, with a shaded region indicating uncertainty/error in the observed profile of $V_{exp}/V_{LE}$ = ($z$) which is estimated by using the method of error propagation as given below:

\begin{align} \label{equ:vexp/vle_error}
    \Delta z = \sqrt{ \left(\frac{\partial z}{\partial \kappa}\right)^2\Delta k^2 + \left(\frac{\partial z}{\partial h}\right)^2\Delta h^2 + \left( \frac{\partial z}{\partial (d \kappa/dh)}\right)^2 \Delta \left(\frac{d \kappa}{dh}\right)^2}
\end{align}

$$\mathrm{where;}~\frac{\partial z}{\partial \kappa} = \frac{1}{(1+\kappa)^2} - \frac{2h}{(1+\kappa)^3}\frac{dk}{dh}$$

$$\frac{\partial z}{\partial h} = \frac{1}{(1+\kappa)^2} \frac{dk}{dh}$$

$$\frac{\partial z}{\partial (d \kappa/dh)} = \frac{h}{(1+\kappa)^2}$$

For all selected events, we further use the derivative of both proposed functions as given in Equation~\ref{equ:func1diff} and Equation~\ref{equ:func2diff} in Equation~\ref{equ:vexp/vle} to do weighted least-squares fit the observed/reference (orange curve) in Figure~\ref{fig:vexp_vle_funct}. The blue and green curves correspond to the fits obtained using the functions $f_1'(h,\kappa)$ and $f_2'(h,\kappa)$, respectively. We note that using derivative of both provide a good fit to the observations. However, $f_1'(h,\kappa)$ yields estimates that are closer to the observed values, as indicated by the smaller errors.

\begin{align} \label{equ:func1diff}
    f_1'(h,\kappa) = &\frac{A (h^{n-1}e^{-k_1 h})}{a^n(1 + e^{-k_1 h})}\left[ n - \frac{k_1 h}{1 + e^{-k_1 h}}\right] + \nonumber \\
    &\frac{B(e^{-k_2h})}{(1 + e^{-k_1 h})(1 + e^{-k_2 h})} \left[ \frac{k_1e^{-k_1h}}{1 + e^{-k_1 h}} - \frac{k_2}{1 + e^{-k_2 h}}\right] \nonumber \\  & + \frac{C(h^{-(m+1)})}{b^{-m}(1 + e^{-k_2h})} \left[\frac{hk_2e^{-k_2h}}{1 + e^{-k_2 h}} - m \right]
\end{align}

\begin{equation} \label{equ:func2diff}
    f_2'(h,\kappa) = f_2(h, \kappa) \left[ \frac{n}{h} - \frac{\beta \gamma h^{\gamma -1}}{a^{\gamma}} - \frac{\delta}{b}\left( 1 + \frac{h}{b}\right)^{-1}\right]
\end{equation}

The last three columns of Table~\ref{tab:tab_2} present the ratio of radial expansion speed to leading-edge speed ($V_{\rm exp}/V_{\rm LE}$) at the last tracked height and 1 AU. The value at 1 AU is listed from using Equation~\ref{equ:vexp/vle} and also directly from in situ measurements of speeds of MCs. At coronal heights, this ratio is derived from the GCS model (with uncertainties propagated from speed errors discussed in Section~\ref{sec:errorsgcs}) and shows generally higher values for fast CMEs, suggesting more efficient expansion during early propagation. The values estimated from Equation~\ref{equ:vexp/vle} are broadly consistent with the those from in situ measurements within uncertainties (corrected for spacecraft angular offset described in Section~\ref{sec:errspdratio}), supporting the validity of the adopted approach. We note that the in situ values (from both approach as listed in last two columns of Table~\ref{tab:tab_2}) at 1 AU are systematically lower, indicating a reduction in relative expansion during interplanetary propagation. Overall, the comparison highlights a transition from stronger expansion in the corona to comparatively weaker expansion at 1 AU, with notable differences between fast and slow CMEs.

From Table~\ref{tab:tab_2} and Figure~\ref{fig:vexp_vle_funct}, we examine the evolution of the radial expansion-to-leading-edge speed ratio ($V_{\rm exp}/V_{\rm LE}$) from the final tracked coronal height ($\sim15\,R_{\odot}$) to that at 1 AU from Equation~\ref{equ:vexp/vle}. For the fast CMEs, the ratio decreases from $18.7\pm4.5$\% to $6.7\pm2.4$\% for the 2011 Jun 02 event ($\sim64\%$ reduction), from $38.9\pm7$\% to $21.9\pm1.3$\% for the 2012 Jul 12 event ($\sim44\%$ reduction), from $28.6\pm9$\% to $6.5\pm1.6$\% for the 2013 Apr 11 event ($\sim77\%$ reduction), and from $37.5\pm7.8$\% to $2.3\pm5.7$\% for the 2013 Sep 29 event ($\sim94\%$ reduction).

Similarly, for the slow CMEs, the ratio decreases from $18.7\pm3.6$\% to $1.3\pm1.1$\% for the 2008 Dec 12 event ($\sim93\%$ reduction), from $27.0\pm4.3$\% to $14.2\pm4.5$\% for the 2010 Apr 03 event ($\sim47\%$ reduction), from $21.2\pm4$\% to $2.3\pm0.3$\% for the 2010 Apr 08 event ($\sim89\%$ reduction), and from $30\pm5$\% to $2.8\pm3.5$\% for the 2011 Sep 13 event ($\sim91\%$ reduction).

Overall, both fast and slow CMEs exhibit a substantial decline in $V_{\rm exp}/V_{\rm LE}$ with increasing heliocentric distance, with reductions ranging from $\sim44\%$ to $\sim94\%$, indicating a systematic decrease in the efficiency of radial expansion during interplanetary propagation. On average, the decrease in $V_{\rm exp}/V_{\rm LE}$ is $\sim70\%$ for fast CMEs and $\sim80\%$ for slow CMEs when all events are considered. From our analysis, we identified one exceptional case in each category deviating from the general trend of the group. Excluding the exceptional cases (2013 Sep 29 for fast CMEs and 2010 Apr 03 for slow CMEs), the average decrease becomes $\sim60\%$ for fast CMEs and $\sim90\%$ for slow CMEs, indicating a more consistent and pronounced decline for slow CMEs \citep{Cremades2020}. It is possible that the expansion of slow CMEs diminishes more rapidly relative to their leading-edge motion compared to fast CMEs. This likely reflects stronger confinement of slow CMEs by the ambient solar wind and a greater ability of fast CMEs to maintain expansion even at 1 AU due to their higher internal pressure.

\subsection{Uncertainties and Corrections in the Expansion-to-Leading-Edge Speed Ratio of CMEs}{\label{sec:errspdratio}}

The uncertainties in $V_{\mathrm{exp}}/V_{LE}$ derived using Equation~\ref{equ:vexp/vle} are computed using Equation~\ref{equ:vexp/vle_error}, taking into account the percentage correction in the $h$ and $\kappa$. These uncertainties are shown with the shaded region over the orange curve in Figure~\ref{fig:vexp_vle_funct}. The errors in these speed estimate ratios over the GCS model fitted heights and over the distances up to 30 $R_{\odot}$ are computed using the percentage uncertainties in $h$ and $\kappa$ as discussed in Section~\ref{sec:errorsgcs} and listed in Table~\ref{tab:GCS_para}. The error in these speed ratios (from Equation~\ref{equ:vexp/vle}) at 1 AU is based on the percentage uncertainties in $\kappa$, by accounting for the spacecraft offset from the MC propagation direction, as described in Section~\ref{sec:erraspectmc}. This fractional uncertainty in $\kappa$  and uncertainty in $h$ (same as in GCS model) is adopted to estimate the uncertainties back up to 30 $R_{\odot}$.

Further, the correction applied to the ratio $V_{\mathrm{exp}}/V_{LE}$ based on in situ measurements of MCs at 1 AU is also listed in Table~\ref{tab:tab_2}. This correction accounts for the angular offset of the spacecraft from the MC propagation direction, which leads to an underestimation of the translational (center) component of the speed, while the expansion component remains unaffected. Accordingly, the true leading-edge speed is estimated as
\begin{equation}
(V_l)_{\mathrm{t}} = \frac{V_l - V_{\mathrm{exp}}}{\cos\Delta} + V_{\mathrm{exp}},
\end{equation}

where $V_l$ is the in situ measured leading-edge speed and $\Delta$ is the angular separation between the spacecraft and the MC propagation direction. Based on the estimated values of $(V_l)_{\mathrm{t}}$, the correction in terms of absolute magnitudes and corresponding percentage change (increase) in the $V_l$ at 1 AU are provided in Table~\ref{tab:del_corr_R_vle}. It is evident that the measured $V_l$ values are underestimated and require correction to match $(V_l)_{\mathrm{t}}$. Using these corrected estimates, we derive corrected values of $V_{\mathrm{exp}}/V_{LE}$ (Table~\ref{tab:tab_2}). The corrected speed ratios are systematically smaller than those obtained directly from the in situ measurements. The negative percentage values in the second-last column of the table indicate the extent to which the measured ratios must be reduced. We note that additional uncertainties may affect the estimates obtained from Equation~\ref{equ:vexp/vle}, since the $V_{\mathrm{exp}}$ values derived from in situ observations do not necessarily represent the instantaneous expansion speed at the time of CME leading-edge arrival \citep{Agarwal2024}.

By comparing CME LE speeds near the Sun and at 1 AU, our results indicate that the relative contribution of expansion decreases more significantly for slow CMEs than for fast CMEs. While the three-phase evolution of CME LE kinematics has been reported in earlier studies \citep{Zhang2006}, our findings emphasize that the underlying physical processes governing this evolution remain poorly understood. The assumptions adopted in this analysis, and the resulting limitations, are discussed below.

\section{Results and Discussion} \label{sec:resdis}

We investigate the evolution of CME aspect ratio ($\kappa$), radial size, and expansion dynamics from the low–middle corona to 1 AU for 8 Earth-directed, non-interacting CMEs from the \textit{STEREO} era, whose associated magnetic clouds (MCs) were observed by the \textit{Wind} spacecraft. The analysis combines GCS-model-derived measurements in the corona, inferred behavior at intermediate heliocentric distances, and corrected in situ estimates at 1 AU.  We note that the aspect ratio increases by $\sim110\%$ for fast CMEs and $\sim60\%$ for slow CMEs over the tracked coronal heights, which have implications for assumptions of self-similar expansion. This indicates stronger early expansion in fast CMEs, likely driven by their enhanced post-eruption magnetic fields \citep{Morosan2022}.

We compare our results with earlier studies on the CME aspect ratio. \citet{Patsourakos2010} and \citet{Veronig2018} defined the aspect ratio as the ratio of CME center height to its radius (reciprocal of our definition), while \citet{Krall2001} used the ratio of center height to diameter (half the reciprocal of our definition). Using our definition, \citet{Patsourakos2010} reported an aspect ratio of $\sim0.5$ at $\sim2$--3.5 $R_\odot$, which is higher than our values ($\sim0.27$ for fast and $\sim0.22$ for slow CMEs), whereas \citet{Veronig2018} obtained $\sim0.25$, comparable to our estimates. At larger heights, \citet{Krall2001} reported $\sim0.2$ at $\sim25\,R_\odot$, lower than our values ($\sim0.46$ for fast and $\sim0.33$ for slow CMEs). These differences likely arise from variations in fitting methods and sample size.

We note that on average, the MCs associated with fast CMEs are larger and exhibit higher aspect ratios at 1 AU than those associated with slow CMEs. The mean measured radii for slow and fast events at 1 AU are $\sim20.2$ and $\sim31.5\,R_{\odot}$, increasing to corrected true radii of $\sim24$ and $\sim35.4\,R_{\odot}$, respectively. Similarly, the mean in situ measured aspect ratios of $\sim0.11$ (slow) and $\sim0.18$ (fast) become $\sim0.12$ and $\sim0.21$ after correction. These results suggest that fast CMEs tend to evolve into larger and more radially extended structures by 1 AU. The average radius of MCs in our study is smaller than those reported in earlier studies \citet{Liu2005,Mishra2021a,Zhuang2023}. This also implies that, for comparable values of geoeffective parameters, faster CMEs may result in longer impact durations of geomagnetic disturbances than slower CMEs.

Using the corrected values of $\kappa$ at 1 AU, the aspect ratio decreases from the last tracked coronal height by $\sim35$--75\% for fast CMEs (average $\sim55\%$) and $\sim25$--75\% for slow CMEs (average $\sim60\%$). Excluding the exceptional cases (2013 Sep 29 for fast CMEs and 2010 Apr 03 for slow CMEs), the average decrease in the corrected aspect ratio becomes $\sim45\%$ for fast CMEs and $\sim75\%$ for slow CMEs. This shows a substantial reduction in $\kappa$, confirming a genuine decline in radial expansion efficiency during interplanetary propagation.

Our results show that CME aspect ratio does not remain constant; instead, it exhibits a three-phase evolution with heliocentric distance. In the low-middle corona ($\lesssim10$--$15\,R_{\odot}$), $\kappa$ shows a \textit{rise phase} where it increases rapidly, indicating strong radial expansion, with a more pronounced rise for fast CMEs. At intermediate heights ($\sim15$--$30\,R_{\odot}$), $\kappa$ reaches a plateau \textit{(saturation phase)}, reflecting a regime of regulated expansion. Beyond this, $\kappa$ shows a \textit{decay phase}, based on our corrected in situ measurements, indicating a reduced efficiency of radial expansion during interplanetary propagation. While both fast and slow CMEs follow this trend, fast CMEs attain higher peak $\kappa$ values and show greater variability, whereas slow CMEs exhibit a more gradual rise and a comparatively faster decline. This also suggests that CMEs do not maintain strict self-similar expansion from the Sun to 1 AU, with implications for the estimation of CME arrival time and impact duration at Earth.

For all selected CMEs, the decrease in $\kappa$ from $\sim30\,R_{\odot}$ to its value at 1 AU is described by a power-law relation. On average, fast and slow CMEs follow $\kappa \propto h^{-0.51}$ and $\kappa \propto h^{-0.63}$, respectively, indicating a steeper decline for slow CMEs. These indices may carry uncertainties due to the assumption of a fixed reference height ($\sim30\,R_{\odot}$) where $\kappa$ is taken to be constant, which may vary between events. This can be further constrained using multipoint in situ observations or heliospheric imaging with the GCS model.

The ratio of radial expansion speed to leading-edge speed ($V_{\rm exp}/V_{\rm LE}$) serves as a direct measure of CME expansion efficiency. Derived from the evolution of $\kappa$ and independently estimated from in situ measurements, this ratio shows a substantial decrease from near the Sun to 1 AU. On average, the decrease is $\sim70\%$ for fast CMEs and $\sim80\%$ for slow CMEs. Excluding the exceptional cases (2013 Sep 29 for fast CMEs and 2010 Apr 03 for slow CMEs), the average decrease becomes $\sim60\%$ for fast CMEs and $\sim90\%$ for slow CMEs, indicating a more consistent and pronounced decline for slow CMEs.

To model the evolution of $\kappa(h)$, we employ two functions, $f_1(h,\kappa)$ and $f_2(h,\kappa)$ (Equations~\ref{equ:func1} and \ref{equ:func2}), designed to capture the rise, saturation, and decline phases. The reference $\kappa$ profile combines GCS-derived coronal values with the inferred evolution constrained by corrected in situ measurements at 1 AU, and the fits are obtained using a weighted least-squares approach. Both functions reproduce the observed evolution and the derived $V_{\rm exp}/V_{\rm LE}$ profiles well; however, $f_1(h,\kappa)$ consistently shows smaller deviations (Figure~\ref{fig:kap_funct_fast_slow}), indicating that an explicit phase-based formulation provides a slightly better description than the continuous form of $f_2(h,\kappa)$.

The observed three-phase evolution of $\kappa$ and $V_{\rm exp}/V_{\rm LE}$ reflects a transition in the dominant physical processes governing CME expansion. This transition from the rise phase to the decline phase via the saturation phase represents a shift from magnetically dominated expansion in the corona to a regime increasingly controlled by the heliospheric environment. The consistency between the decrease in $\kappa$ and the reduction in $V_{\rm exp}/V_{\rm LE}$ provides strong kinematic evidence for a evolving expansion-propagation dynamics during interplanetary propagation of CMEs/MCs.

Several MHD models use the 3D kinematics of CMEs at 0.1 AU as initial boundary conditions; however, they often do not  incorporate the evolution of CME aspect ratio in their simulations \citep{Odstrcil2003,Pomoell2018,Mayank2024}. Also, aspect-ratio-related effects may be incorporated in the HUXt model to examine the performance of the model \citep{Barnard2022,Owens2025}. Based on our analysis, the variation in CME aspect ratio reflects the evolving expansion of CMEs, which, if not properly accounted for, may affect estimates of CME radial size, CME duration, arrival time of CME substructures, and embedded magnetic flux. Although the interplay between CME aspect ratio and external factors such as heliospheric magnetic fields, solar wind structures, and CME--CME interactions remains poorly understood, our results provide new observational constraints for investigating the role of evolving CME aspect ratio and improving future space-weather models.

In this study, we combine multipoint remote observations up to $\sim15\,R_{\odot}$ with in situ measurements at 1 AU to investigate the evolution of CME aspect ratio. Remote estimates may overestimate the CME radius if the bright leading edge and compressed material are included in GCS fitting \citep{Vourlidas2013,Temmer2022}, while the estimates from in situ measurements may be underestimated if the spacecraft angular offset is not properly accounted. Although the selected CMEs/MCs are Earth-directed and in situ measurements are corrected, uncertainties in aspect ratio may still arise from idealized geometrical considerations \citep{Demoulin2009,Wang2015}, breakdown in coherency in MC structure in the interplanetary medium \citep{Owens2017,Agarwal2025}, and possible deflection, rotation, or interaction with the ambient solar wind or other large-scale solar wind structure \citep{Wang2004,Kay2015,Mishra2016}. Magnetic erosion due to reconnection may further reduce the MC radial size \citep{Owens2006,Ruffenach2012}. Despite these limitations, our results demonstrate a three-phase evolution of CME aspect ratio and expansion dynamics in the heliosphere. Future studies using multipoint in situ observations and heliospheric imaging will help further constrain this evolution and improve our understanding of CME aspect ratio and its implications for space weather.

\section*{Acknowledgements}

We thank the teams of the STEREO/COR, SOHO/LASCO (C2 and C3), and Wind spacecraft for making their observational data publicly available. We also thank the referee for the constructive comments that helped improve the manuscript.

\section*{Data Availability}
The data sets used and/or analyzed during this study are available from the corresponding author upon reasonable request.




\bsp	
\label{lastpage}
\end{document}